\documentclass[aps,pra,showpacs,twocolumn,amsmath,amssymb]{revtex4}
\usepackage{graphicx}
\usepackage{epsfig}
\usepackage{color}
\usepackage{bm}
\usepackage{amssymb}
\usepackage{amsmath}
\usepackage{latexsym}
\usepackage{longtable}
\definecolor{highlight}{rgb}{1.0,0.0,0.0}
\newcommand{\ind}[1]{_{#1}}    
\newcommand{\thick}{d}   
\newcommand{\vc}[1]{\mbox{\boldmath $#1$}} 
\newcommand{\ekspon}[1]{{\mathrm e}^{\textstyle #1}} 
\newcommand{\deo}{\varkappa}   
\newcommand{\iu}{{\mathrm i}}       
\newcommand{\1}{1}
\newcommand{\2}{2}

\newcommand{\gvvec}{\vc{V}_{\nu}} 
\newcommand{\gvval}{V_{\nu}} 
\newcommand{\dev}{y}
\newcommand{\whh}{w_{\ind{H}}}   
\newcommand{\dhkl}{d_{\ind{H}}}   
\newcommand{\whhs}{w_{\ind{H}}^{\ind{{\mathrm {(s)}}}}}   
\newcommand{\extleng}{\bar{\Lambda}_{\ind{H}}}
\newcommand{\extlengs}{{\extleng ^{\ind{{\mathrm {\!(s)}}}}}}

\newcommand{\OmegaB}{\Omega_{\ind{\mathrm B}}}
\newcommand{\timex}{\mathcal{T}_{\ind{\Lambda}}}
\newcommand{\timexo}{\mathcal{T}_{\ind{0}}}
\newcommand{\omegaa}{\whh\omegab}
\newcommand{\exttimes}{\mathcal{T}_{\ind{\Lambda}}^{\ind{{\mathrm {\!(s)}}}}}
\newcommand{\thetax}{\tilde{\theta}}
\newcommand{\hvar}{{\mathcal{H}}}
\newcommand{\gammah}{\gamma_{\ind{H}}}   
\newcommand{\gammaht}{\tilde{\gamma}_{\ind{H}}}   
\newcommand{\gammao}{\gamma_{\ind{0}}}   
\newcommand{\Deltah}{\Delta_{\ind{H}}}

\newcommand{\Deltaeh}{\Delta E_{\ind{H}}}
\newcommand{\Deltaeo}{\Delta E_{\ind{0}}}
\newcommand{\tauh}{\tau_{\ind{H}}}
\newcommand{\tauo}{\tau_{\ind{0}}}
\newcommand{\taux}{\tau_{\ind{\hvar}}}
\newcommand{\tauxo}{\xi_{\ind{\hvar}}}
\newcommand{\tauho}{\xi_{\ind{H}}}

\newcommand{\tauoo}{\xi_{\ind{0}}}
\newcommand{\psih}{\psi_{\ind{H}}}
\newcommand{\psio}{\psi_{\ind{0}}}

\newcommand{\omegab}{\omega}
\newcommand{\omegabc}{\omegab}

\newcommand{\wshift}{v}
\newcommand{\uc}{u}
\newcommand{\ucvco}{\vc{\hat{\uc}}_{\ind{0}}}
\newcommand{\ucvch}{\vc{\hat{\uc}}_{\ind{H}}}
\newcommand{\ucvcx}{\vc{\hat{\uc}}_{\ind{\hvar}}}
\newcommand{\wcvco}{\vc{\hat{\wshift}}_{\ind{0}}}
\newcommand{\wcvch}{\vc{\hat{\wshift}}_{\ind{H}}}
\newcommand{\wcvcx}{\vc{\hat{\wshift}}_{\ind{\hvar}}}
\newcommand{\wco}{\wshift_{\ind{0}}}
\newcommand{\wch}{\wshift_{\ind{H}}}
\newcommand{\wcx}{\wshift_{\ind{\hvar}}}

\newcommand{\ebr}{E_{\ind{H}}}

\newcommand{\thetac}{\theta}

\newcommand{\omegao}{\omega_{\ind{0}}}
\newcommand{\ko}{K_{\ind{0}}}
\newcommand{\kovccrystal}{\vc{k}_{\ind{\! 0}}}
\newcommand{\kovc}{\vc{K}_{\ind{\! 0}}}
\newcommand{\khvc}{\vc{K}_{\ind{\! H}}}
\newcommand{\kxvc}{\vc{K}_{\ind{\! \hvar}}}
\newcommand{\rhvc}{\vc{r}_{\ind{\! H}}}
\newcommand{\rovc}{\vc{r}_{\ind{\! 0}}}
\newcommand{\rxvc}{\vc{r}_{\ind{\! \hvar}}}
\newcommand{\rvc}{\vc{r}}

\newcommand{\refl}{R}
\newcommand{\fieldin}{\vc{\mathcal{D}}}
\newcommand{\fieldinmono}{\vc{\mathcal{D}}^{\mathrm {(m)}}}
\newcommand{\fieldout}{\vc{\mathcal{E}}}
\newcommand{\fieldoutmono}{\vc{\mathcal{E}}^{\mathrm {(m)}}}
\newcommand{\fieldinc}{\vc{\cal E}_{\ind{\mathrm i}}}
\newcommand{\incamp}{\vc{\cal E}_{\ind{\mathrm i}}}
\newcommand{\cnstc}{\mathcal{C}}

\newcommand{\cnsta}{\mathcal{A}}
\newcommand{\cnstd}{\mathcal{G}}
\newcommand{\thicknessduration}{\mathcal{T}_{\ind{d}}}
\newcommand{\dirate}{D}   
\newcommand{\sgn}[1]{{\mathrm {sgn}}\left\{ {#1} \right\}}
\newcommand{\deltabl}{\delta_{\ind{BL}}}

\begin{document}
\title{Spatiotemporal Response of  Crystals in  X-ray Bragg Diffraction}
\author{Yuri Shvyd'ko}\email{shvydko@aps.anl.gov}   \affiliation{Advanced Photon Source,  Argonne National Laboratory, Argonne, Illinois 60439, USA}
\author{Ryan Lindberg}\email{lindberg@aps.anl.gov}  \affiliation{Advanced Photon Source,  Argonne National Laboratory, Argonne, Illinois 60439, USA}
\date{\today}
\begin{abstract} 
  The spatiotemporal response of crystals in x-ray Bragg diffraction
  resulting from excitation by an ultra-short, laterally confined
  x-ray pulse is studied theoretically. The theory presents an
  extension of the analysis in symmetric reflection geometry
  \cite{LS12} to the generic case, which includes Bragg diffraction
  both in reflection (Bragg) and transmission (Laue) asymmetric
  scattering geometries.

  The spatiotemporal response is presented as a product of a
  crystal-intrinsic plane wave spatiotemporal response function and an
  envelope function defined by the crystal-independent transverse
  profile of the incident beam and the scattering geometry.  The
  diffracted wavefields exhibit amplitude modulation perpendicular to
  the propagation direction due to both angular dispersion and the
  dispersion due to Bragg's law.  The characteristic measure of the
  spatiotemporal response is expressed in terms of a few parameters:
  the extinction length, crystal thickness, Bragg angle, asymmetry
  angle, and the speed of light.

  Applications to self-seeding of hard x-ray free electron lasers are
  discussed, with particular emphasis on the relative advantages of
  using either the Bragg or Laue scattering geometries.  Intensity
  front inclination in asymmetric diffraction can be used to make
  snapshots of ultra-fast processes with femtosecond resolution.
\end{abstract}
%
\pacs{41.50.+h,42.25.-p, 61.05.cp, 07.85.Nc}
%
%
\maketitle

\section{Introduction}

The spatiotemporal response from crystals in symmetric x-ray Bragg
diffraction in reflection (Bragg) geometry has been studied in our
recent publication \cite{LS12}. Here, we extend the analysis to
the generic case of asymmetric Bragg diffraction, both
in reflection (Bragg) and transmission (Laue)
geometries. ``Asymmetric'' means that the Bragg reflecting atomic
planes are not parallel to the crystal surface.


Understanding the time dependence of x-ray Bragg diffraction in
crystals has attracted much attention since the 1990s, as the advent
of ultra-fast (femtosecond short) x-rays pulses become a close
reality.  The temporal and spatial dependence of diffraction was first
calculated using the time dependent Takagi-Taupin equations.  In
particular, an analytical solution for the Bragg reflected wave from
an infinitely thick crystal was derived by Chukhovskii and F\"orster
\cite{ChFo95}.  Numeric calculations of the time dependence of Bragg
diffraction from a crystal heated by a laser pulse was performed in
\cite{Wark99}.  Calculations of the time dependence by Fourier
transforming the known monochromatic plane-wave solutions from the
classical dynamical theory
\cite{Ewald17,Laue31,Zach,Laue60,BC64,Pinsker,Pinsker82,Authier} have
been considered in several publications
\cite{SZM01,SZM01-2,Siddons01,Graeff02,MG03,Siddons04,Shvydko-SB,Graeff04}.
In particular, Shastri et al. \cite{SZM01,SZM01-2} performed numerical
calculations, which have revealed signature features of time
dependences of Bragg diffraction from crystals both in the Bragg-case
and in the Laue-case geometries.  Graeff and Malgrange
\cite{Graeff02,MG03} obtained analytical solutions for the time
dependence of Bragg diffraction in the Laue geometry, with the
refraction effects at the crystal exit surface taken into
account. Bushuev \cite{Bushuev08} used Fourier transformation of the
plane-wave solutions both in the frequency and momentum space, with
the second order corrections included to more accurately account for
the refraction effects, to obtain solutions in time and space and
analyzed specific cases using numeric calculations.

The present paper is focused on the development of the theory and on
the analysis of the spatiotemporal response of crystals in Bragg
diffraction to the excitation by an ultra short in time and spatially
confined x-ray pulse in the general case of asymmetric reflection
(Bragg) and transmission (Laue) scattering geometries.  The primary
goal of the present study is to understand the general phenomenology
of the spatiotemporal response by uncovering the dominant underlying
physics and identifying the key physical parameters that determine the
characteristic time and space scales involved.  For this purpose, we
derive comprehensive solutions that can be written in the general case
as a product of two independent envelope functions: the first is a
spatiotemporal plane-wave response function that depends only on the
crystal and scattering geometry, while the second is an envelope that
is specific to the initial conditions of the incident field.  We
derive analytical solutions for the response functions under several
representative conditions, which clearly identifies the key physical
parameters and makes possible a relatively simple interpretation of
the general solution.

When an ultra-short x-ray pulse instantaneously excites a perfect
crystal, the output field is delayed and spread in time.  The
underlying reason behind this phenomenon is that each frequency
component excites a monochromatic eigen-wavefield in the crystal that
propagates along its own direction with its associated group velocity.
As a consequence, the time response is intrinsically connected to the
lateral spatial distribution of x-rays leaving the crystal upon Bragg
diffraction both in the reflection or forward directions.

The paper is organized as follows.  Comprehensive solutions for the
spatiotemporal dependences of Bragg diffraction in reflection (Bragg)
and transmission (Laue) asymmetric geometries are derived in
Sec.~\ref{theory}.  In particular, in Sec.~\ref{boundless} the
solutions are derived for incident ultra-short x-ray pulses with an
unbounded plane wave front, and in Sec.~\ref{pencil-beam} for incident
ultra-short x-ray pulses with a bounded wavefront. The solution for
the bounded wavefront is a product of a crystal-intrinsic and
geometry-specific plane wave spatiotemporal response function and an
envelope function defined by the crystal-independent transverse
profile of the incident beam and the scattering geometry.  The
response functions in the asymmetric Bragg geometry are derived
analytically in Sect.~\ref{response-bragg} in the approximation of a
non-absorbing and thick $\thick \gg \extlengs$ crystal (the crystal
thickness $\thick$ being much larger than the characteristic
extinction length of Bragg diffraction $\extlengs$ to be more
precisely defined below). The response functions in the asymmetric
Laue geometry are derived in analytical form in
Sect.~\ref{response-laue} for a non-absorbing crystal of arbitrary
thickness. Applications of the theory for self-seeding of XFELs and
for ultra-fast time measurements are discussed in
Sec.~\ref{applications}.

\section{Comprehensive Solutions for Spatiotemporal Crystal Response}
\label{theory}

We study here the spatiotemporal dependence of Bragg diffraction of
ultra-short, laterally bound x-ray pulses from a system of parallel
atomic planes in a flat crystal plate.  Generic solutions are derived
in three consecutive steps.  First, well-known solutions of the
dynamical theory of x-ray Bragg diffraction in crystals
\cite{Ewald17,Laue31,Zach,Laue60,BC64,Pinsker,Pinsker82,Authier,Shvydko-SB}
for incident monochromatic plane waves are briefly summarized in
Sec.~\eqref{monochromatic}. In Sec.~\ref{boundless}, we derive
solutions for an initially ultra-short incident pulse with boundless
plane wavefront.  Finally, solutions are obtained in
Sec.~\ref{pencil-beam} for an ultra-short incident pulse with confined
wavefront.

\subsection{Monochromatic  Plane Wave  Solutions} 
\label{monochromatic}

One of most fundamental results of the dynamical theory of x-ray
diffraction in perfect crystals is the concept of monochromatic eigen
wavefields in crystals introduced by Ewald almost 100 years ago
\cite{Ewald17}. A similar concept in the electron theory of solids was
introduced later by Bloch in 1928, which are therefore generally known
as Bloch waves. In the simplest case, an incident monochromatic plane
wave $\incamp\, \exp(\iu\kovc\vc{r}-\iu\omegao t)$, with a frequency
$\omegao$ and wavevector $\kovc=({\omegao}/{c})\ucvco$, propagating
along the optical axis $\ucvco$, excites in the crystal a wavefield
\begin{equation}
\vc{\fieldin}(\vc{r},t)=\incamp\, \exp(-\iu\omegao t) \sum_{\ind{H}}\,R_{\ind{0H}}\,\exp\left[\iu(\kovc+\vc{H})\vc{r}\right]
\label{wavefield}
\end{equation}
which is a sum of plane wave components with wave vectors
$\kovc+\vc{H}$. In the general case this sum involves all the
reciprocal crystal lattice vectors $\vc{H}$ of the crystal. For each
$\vc{H}$ there is a set of parallel atomic planes in the crystal
perpendicular to $\vc{H}$ with an interplanar distance $\dhkl=2\pi/H$
which actually composes the grating on which x-rays diffract.

In the following we will consider the so-called two-wave case, where
only two plane wavefield components are taken into account: the wave
associated with forward Bragg diffraction $\vc{H}=0$, and one Bragg
diffraction component with nonzero $\vc{H}$, for which
$|\kovc+\vc{H}|\simeq |\kovc|\equiv \ko$, and for which therefore the
relative difference
\begin{equation}
\alpha = \frac{(\kovc+\vc{H})^2-\ko^2}{\ko^2} = \frac{2\kovc\vc{H}+\vc{H}^2}{\ko^2}
\label{pro017}
\end{equation}
has a very small magnitude. In particular, if $\alpha=0$ we obtain Bragg's law $2\kovc\vc{H}+\vc{H}^2=0$, which can be also written as
\begin{equation}
2\ko\sin\theta=H.
\label{bragg}
\end{equation}
Here, $\theta$ is the  glancing angle of incidence to the
atomic planes, which equals the angle between $\kovc$ and the atomic
planes such that $\kovc\vc{H}=-\ko H \sin\theta$.
The quantity $\alpha$ \eqref{pro017} is an important parameter of the theory known as the deviation parameter, since it represents the deviation from Bragg's law.

\begin{figure*}[t!]
\setlength{\unitlength}{\textwidth}
\begin{picture}(1,0.49)(0,0)
\put(0.0,0.00){\includegraphics[width=0.49\textwidth]{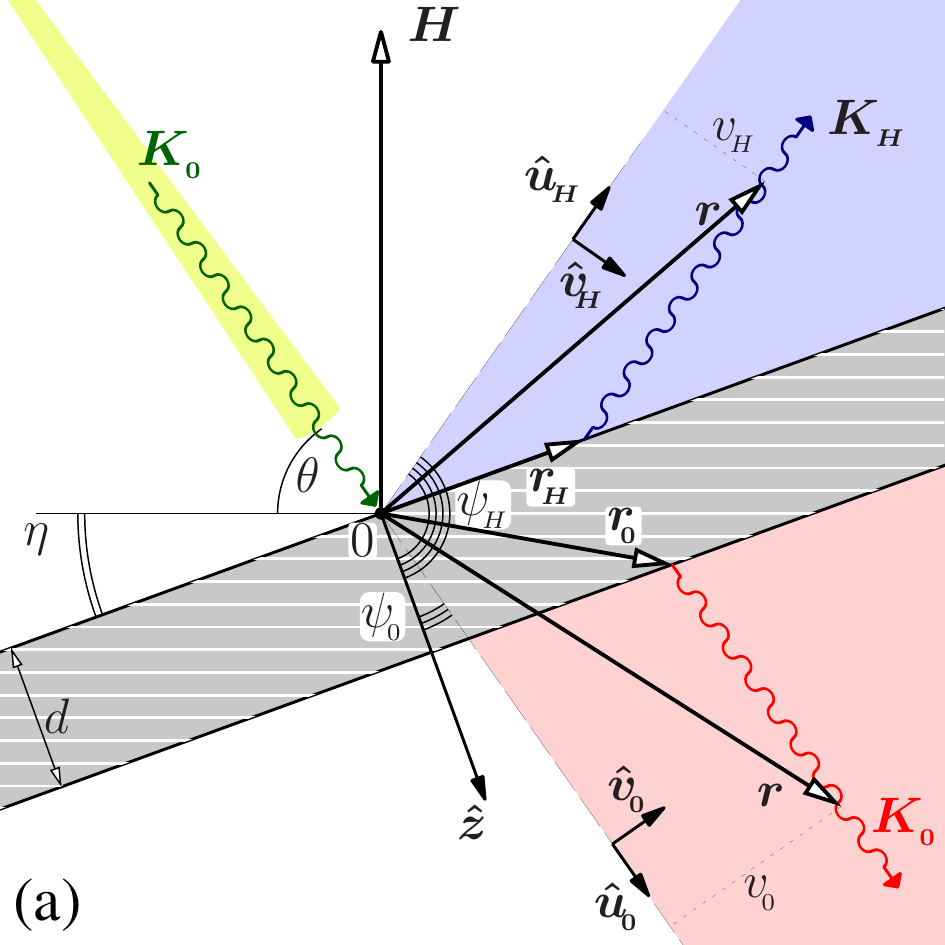}}
\put(0.50,0.00){\includegraphics[width=0.49\textwidth]{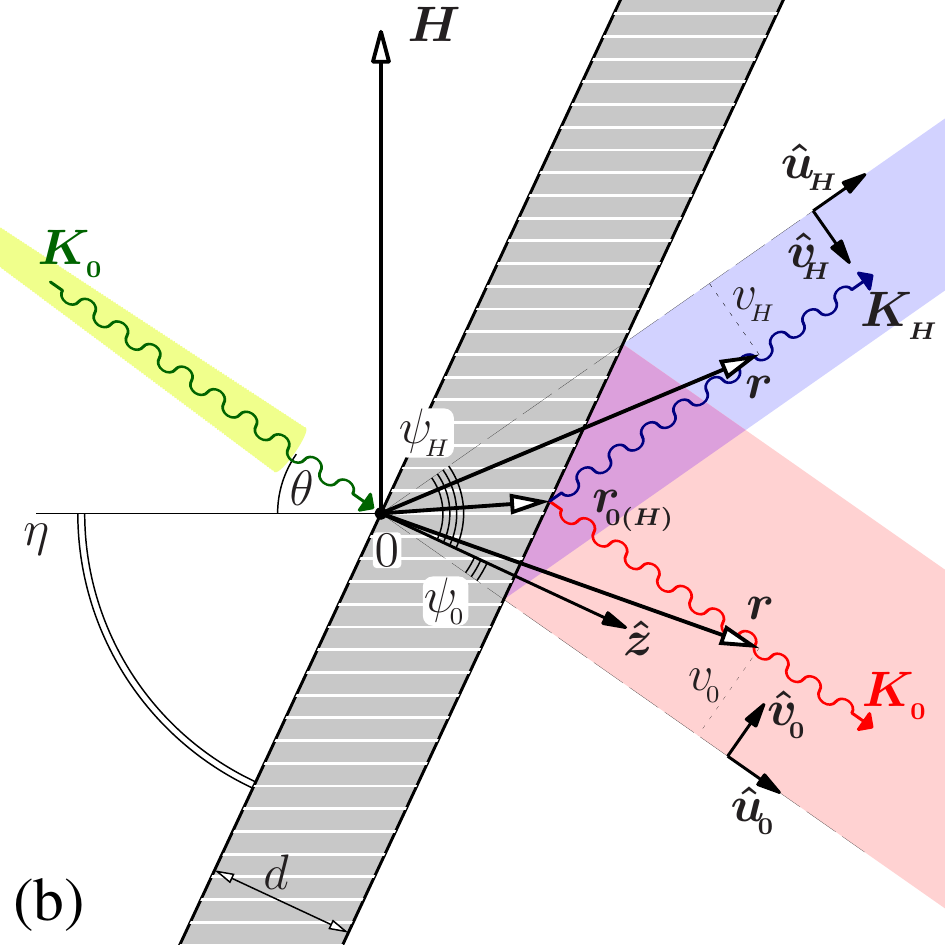}}
\end{picture}
\caption{Schematic presentation of two-beam x-ray Bragg diffraction
  from crystals (a) in the reflection (or Bragg) scattering geometry,
  and (b) in the transmission (or Laue) scattering geometry. The
  glancing angle of incidence to the reflecting atomic planes is
  $\theta$, and the angle between the reflecting planes and the
  crystal surface (the asymmetry angle) is $\eta$.  The propagation
  direction $\ucvch$ of the Bragg reflected beam composes an angle
  $\psih$ with the internal normal $\hat{\vc{z}}$ to the crystal
  surface.  The angle $\psio$ between $\hat{\vc{z}}$ and the direction
  $\ucvco$ of propagation of the incident beam is defined by the
  relationship $\psio+\psih=2\theta$.  For the scattering
  configurations shown in (a) and (b) $\psih=\pi/2+\theta-\eta$, and
  therefore $\psio=\theta+\eta-\pi/2$. The permitted range
    of the asymmetry angle $\eta$ in the Bragg-case geometry is
    $-\theta<\eta<\theta$, while in the Laue-case geometry it is
    $\theta<\eta<\pi-\theta$. Negative $\eta$ in the Bragg-case
    geometry corresponds to a configuration complimentary to that
    shown in (a) with the incident and reflected beams reversed.
    Pink and light blue areas indicate regions where the forward
    diffracted and diffracted beams can propagate.  See text for other
    details and definitions.}
\label{fig002}
\end{figure*}

We take the atomic planes associated with the reciprocal crystal
lattice vector $\vc{H}$ to be oriented at an arbitrary (asymmetry)
angle $\eta$ with respect to the crystal surface, as shown
schematically by white parallel lines in Figs.~\ref{fig002}(a) and
~\ref{fig002}(b).  Figure~\ref{fig002}(a) represents diffraction in
the reflection or Bragg scattering geometry, which is characterized by
the diffracted wavefield exiting the crystal on the same side as the
incident wave, while the forward diffracted wavefield propagates along
the incident wave direction and exits the crystal on the opposite
side.  Figure~\ref{fig002}(b) represents diffraction in the
transmission or Laue scattering geometry, for which both the
diffracted and forward diffracted wavefields exit the crystal from the
surface opposite that of the incident wave.  The crystal surfaces are
defined by the unit normal vector $\vc{\hat{z}}$ internal to the
entrance surface.  We restrict our theory to the case in which $\kovc$
is directed in the plane composed by $\vc{H}$ and $\vc{\hat{z}}$,
hereafter referred to as dispersion plane.

The dynamical theory of x-ray diffraction allows one to calculate for
each incident monochromatic plane wave component $\exp\{-{\mathrm
  i}[\omegao t - \kovc \vc{r}]\}$ both
the monochromatic wavefield of forward Bragg diffraction
\begin{equation}
\fieldinmono_{\ind{\mathrm 0}}(\rovc,t)=\incamp\,\ekspon{-{\mathrm i}[\omegao t  - \kovc \rovc ]}\, \refl_{\ind{00}}(\omegao)
\label{pro0200}
\end{equation}
at any point $\rovc$ on the rear surface of the crystal, and the monochromatic wavefield of Bragg diffraction
\begin{equation}
\fieldinmono_{\ind{\mathrm H}}(\rhvc,t)=\incamp\,\ekspon{-{\mathrm i}[\omegao t - (\kovc+\vc{H}) \rhvc]}\, \refl_{\ind{0H}}(\omegao).
\label{pro0210}
\end{equation}
The Bragg diffracted field \eqref{pro0210} is determined at any point
on the entrance surface $\rhvc$ in the case of Bragg geometry shown in
Fig.~\ref{fig002}(a), or at any point on the rear surface $\rhvc$ for
the Laue geometry shown in Fig.~\ref{fig002}(b). Here,
$\refl_{\ind{0\hvar}}(\omegao)$ are diffraction ($\hvar=H$) or forward
diffraction ($\hvar=0$) crystal amplitudes which are functions not
only of $\omegao$ (whose dependence we explicitly indicate since it is
most relevant to the discussion of the time behavior), but are also
functions of the crystal thickness $d$, the direction and magnitude of
$\kovc$, the asymmetry angle $\eta$, etc.
At this point we are concerned with deriving general expressions for
the spatiotemporal response of the crystal in x-ray Bragg diffraction,
and so do not yet specify these reflection amplitudes; explicit
expressions for the amplitudes $\refl_{\ind{0\hvar}}$ are presented in
Sec.~\ref{response-bragg} and Sec.~\ref{response-laue}.

The in-crystal monochromatic wave field components
$\fieldinmono_{\ind{\mathrm \hvar}}(\rxvc,t)$ given by
Eqs.~\eqref{pro0200} and \eqref{pro0210} can also be used to calculate
the field at any point $\rvc$ outside of the crystal.  Using the
continuity of the wave fields at the crystal-vacuum interface
determined by the extremities of the vectors $\rxvc$, the forward
diffracted and diffracted wave fields in an arbitrary point $\rvc$ in
vacuum can be written as
\begin{equation}
\fieldoutmono_{\ind{\mathrm \hvar}}(\rvc,t)\,=\,\fieldinmono_{\ind{\mathrm \hvar}}(\rxvc,t)\,\, \ekspon{{\mathrm i}\kxvc(\rvc-\rxvc)},
\label{pro025}
\end{equation}
where $\kxvc$ is the wavevector of the forward diffracted ($\hvar=0$)
or diffracted ($\hvar=H$) field in vacuum.  To match phase fronts the
components of the in-crystal wave vectors can differ from the vacuum
wave vectors only by a component along the crystal normal
$\hat{\vc{z}}$. Since we assume that the crystal entrance and exit
surfaces are parallel, this component is zero for the wavevector
$\kovc$, and makes it equivalent to the vacuum wavevector $\kovc$ of the
incident plane wave. However, this component is not zero for the
vacuum wave vector of the diffracted wave. In the general case it can
be written as
\begin{equation} 
\khvc = \kovc +\vc{H} +\Deltah \hat{\vc{z}}.  
\label{pro027}
\end{equation}
The component $\Deltah \hat{\vc{z}}$ can be understood as an
additional momentum transfer due to refraction at the crystal vacuum
interface, and Eq.~\eqref{pro027} as momentum conservation in
scattering from the crystal.  Since Bragg diffraction is an elastic
scattering process and the vacuum is homogeneous, the magnitude of the
vacuum wavevector $\khvc$ of the diffracted wave should be equal to
the the vacuum wavevector of the incident plane wave:
$|\khvc|=|\kovc|\equiv \ko$. From this condition and
Eqs.~\eqref{pro027}, \eqref{pro017} we find:
$\Deltah=\ko(-\gammaht\pm\sqrt{\gammaht^2-\alpha})$ (see
\cite{Shvydko-SB} for details), where
$\gammaht=\hat{\vc{z}}(\kovc+\vc{H})/\ko$. For small
$\alpha$, the additional momentum transfer
can be closely approximated by a Taylor expansion in $\alpha$: 
\begin{equation}
\Deltah = - \ko \frac{\alpha}{2\gammaht}  - \ko\frac{\alpha^{2}}{8\gammaht^3} + \cdots .
\label{pro029}
\end{equation}
Now from Eqs.~\eqref{pro0200}, \eqref{pro0210}, \eqref{pro025}, and
\eqref{pro027} we can write for the monochromatic forward diffracted
and diffracted wavefields
\begin{widetext}
\begin{gather}
\fieldoutmono_{\ind{0}}(\rvc,t)= \incamp \ekspon{-{\mathrm i}[\omegao t - \kovc \rvc ]}\, \refl_{\ind{00}}(\omegao), \label{pro033}\\[0.2cm]
\fieldoutmono_{\ind{H}}(\rvc,t)= \incamp \ekspon{-{\mathrm i}[\omegao t - (\kovc+\vc{H})\, \rvc]}\, \ekspon{{\mathrm i}\Deltah (\rvc-\rhvc)\hat{\vc{z}}}\, \refl_{\ind{0H}}(\omegao). \label{pro0333}
\end{gather}
\end{widetext}
Here $(\rvc-\rhvc)\hat{\vc{z}}$ is the shortest distance from the
observation point $\rvc$ to the crystal surface.  Since $\hat{\vc{z}}$
is perpendicular to the crystal surface, it is actually independent of
$\rhvc$, and $(\rvc-\rhvc)\hat{\vc{z}}=\rvc\hat{\vc{z}}$ if the
extremity of $\rhvc$ is on the entrance surface, and
$(\rvc-\rhvc)\hat{\vc{z}}=\rvc\hat{\vc{z}}- \thick$ if the extremity
of $\rhvc$ is on the rear surface.  Here $\thick$ is the crystal
thickness.

\subsection{Ultra-Short  Incident  Pulse with Boundless Plane Wavefront}
\label{boundless}

To study the spatiotemporal dependence of x-ray diffraction, in the
next step we investigate the response to an initially ultra-short
(instantaneous) x-ray pulse.  We assume that the x-ray pulse
propagates along the direction of the unit vector $\ucvco$ which is in
the dispersion plane built by vectors $\vc{H}$ and $\vc{\hat{z}}$, and
that the propagation direction $\ucvco$ makes a glancing angle of
incidence $\theta$ with respect to the reflecting atomic planes.

The x-ray pulse is ultra-short, has a vector amplitude $\incamp $, and
has an infinite extent in the transverse direction
$\wcvco\perp\ucvco$. In this case, the pulse at time $t$ and spatial
point $\vc{r}$ can be presented by the delta function $\delta(\tau)$
of the argument $\tau\,=\,t-{\ucvco\vc{r}}/{c}$, with $c$ the speed of
light in vacuum. The latter is equivalent to an infinite sum of
monochromatic plane wave components given by
\begin{gather}
  \incamp\ekspon{-{\mathrm i}\omegab\tau}\delta(\tau)\,=\,\incamp\int_{-\infty}^{\infty}\frac{{\mathrm d}\Omega}{2\pi}\,
  \ekspon{-{\mathrm i}{\left(\omegab+\Omega\right)\tau}},\label{pro010}\\
\tau\,=\,t-\frac{\ucvco\vc{r}}{c}, \hspace{0.75cm} \omegab+\Omega\,=\,\omegao, \hspace{0.75cm} \kovc\,=\,\frac{\omegao}{c}\ucvco.
\label{pro012}
\end{gather}
Here, we single out one plane wave component with a frequency
$\omegab$ (we assume $\omegab\gg\Omega$), which we define to satisfy
the condition $\alpha=0$ defined in Eq.~\eqref{pro017}. In other
words, we are selecting out the frequency for which Bragg's law
\eqref{bragg} is fulfilled. For the frequency $\omegab$ Bragg's law
reads $\omega\sin\theta=Hc/2$. With this convention, the deviation
parameter $\alpha$ \eqref{pro017} can be presented as 
\begin{equation}
\alpha\, =\,  -4\,\frac{\Omega}{\omegab}\,\sin^2\theta\,\left(1-2\frac{\Omega}{\omega}+\cdots\right), 
\label{rta050}
\end{equation}
and the additional momentum transfer $\Deltah$ \eqref{pro029} as  
\begin{equation}
\frac{\Deltah}{\omegab/c}\, =  \,\frac{2\sin^2\theta }{\gammah}\,\frac{\Omega}{\omegab}\,\left[1-\frac{\Omega}{\omegab}\left(b+\frac{\sin^2\theta}{\gammah^2}\right)\,+\cdots\right],\label{pro031}
\end{equation}
where
\begin{equation}
b=\frac{\gammao}{\gammah}, \hspace{0.25cm} \gammao=\hat{\vc{z}}\ucvco , \hspace{0.25cm} \gammah=\hat{\vc{z}}\ucvch , \hspace{0.25cm} \ucvch=\ucvco+\frac{\vc{H}}{\omegab/c} \label{asymmetryfactor}
\end{equation}
are the so called asymmetry factor $b$, and direction cosines
$\gammao\equiv\cos\psio$, $\gammah\equiv\cos\psih$. In almost all of
what follows we retain only the terms linear in the small quantity
$\Omega/\omegab$ for the expressions for $\alpha$ \eqref{rta050} and
$\Deltah$ \eqref{pro031}, but we also present a brief description of
the physics of the quadratic terms and how they can be included.

\subsubsection{Linear approximation}

Time $t=0$ is defined hereafter as the moment when the wavefront hits
the point $\vc{r}=0$ on the crystal.  Similar to \eqref{pro010}, the
spatiotemporal response of the crystal in Bragg diffraction
$\fieldout_{\ind{\hvar}}(\rvc,t)$, both for diffracted ($\hvar = H$)
and forward diffracted ($\hvar=0$) components, can be calculated as an
integral (in fact, a Fourier integral)
\begin{equation}
\fieldout_{\ind{\mathrm \hvar}}(\rvc,t)\,=\,\int_{-\infty}^{\infty}\,\frac{{\mathrm d}\Omega}{2\pi}\,\, \fieldoutmono_{\ind{\mathrm \hvar}}(\rvc,t)
\label{pro0125}
\end{equation}
over the monochromatic components \eqref{pro033}-\eqref{pro0333}.  A
similar procedure was also applied in the previous publications
\cite{KAK79,SZM01,SZM01-2,Siddons01,Graeff02,MG03,Siddons04,Shvydko-SB,Graeff04,Bushuev08}.
Using \eqref{pro033}-\eqref{pro0333}, we obtain
\begin{gather}
  \fieldout_{\ind{\hvar}}(\rvc ,t)\,=\,  \incamp\,\ekspon{-{\mathrm i}\omegab \taux }\,G_{\ind{0\hvar}}(\tauxo), \label{pro020} \\
  G_{\ind{0\hvar}}(\tauxo)\,=\,\int_{-\infty}^{\infty}\frac{{\mathrm
      d}\Omega}{2\pi}\,\ekspon{-{\mathrm i}\Omega
    \tauxo}\,\refl_{\ind{0\hvar}}(\omegab+\Omega),
\label{pro022}
\end{gather}
where
\begin{gather}
\taux\,=\,t-\frac{\ucvcx \rvc}{c},\label{pro023}\\
\tauxo\,=\,t-\frac{\ucvco \rvc}{c} - 2 \sin^2\theta \,\frac{(\rvc-\rhvc)\hat{\vc{z}}}{c\gammah} \,\delta_{\ind{\hvar H}}.
 \label{pro024}
\end{gather}
Here, $\delta_{\ind{\hvar H}}$ is the Kronecker delta, which equals one only if $\hvar=H$, otherwise it is zero. 

The plane wave crystal response functions $G_{\ind{0\hvar}}(\tauxo)$
in \eqref{pro022} represent the spatiotemporal dependence of Bragg
diffraction ($\hvar=H$) or forward Bragg diffraction ($\hvar=0$) to
the excitation by a $\delta$-function-short incident radiation pulse
with boundless transverse wavefront.

Two spatiotemporal variables are introduced in
\eqref{pro020}-\eqref{pro024}. The variable $\tauxo$ ($\hvar=0,H$)
\eqref{pro024} is the argument of the response function
\eqref{pro022}. The spatiotemporal variable $\taux$ ($\hvar=0,H$)
\eqref{pro023} is in the argument of the exponential function of
Eq~\eqref{pro020}.

The spatiotemporal variable $\taux$ ($\hvar=0,H$) represents the
difference between the absolute time $t$ and the time ${\ucvcx
  \rvc}/{c}$ the plane wavefront, propagating along the optical axis
$\ucvcx$ from $\rvc=0$, would need to reach an arbitrary point $\rvc$
outside of the crystal, assuming the propagation is in vacuum and the
wavefront is perpendicular to the optical axis $\ucvcx$.  In other
words, $\taux$ is the time delay for the radiation field at point
$\rvc$ we are interested in, as compared to the trivial propagation of
the pulse in vacuum along $\ucvcx$.

\begin{figure}[t!]
\setlength{\unitlength}{\textwidth}
\begin{picture}(1,0.27)(0,0)
\put(0.0,0.00){\includegraphics[width=0.49\textwidth]{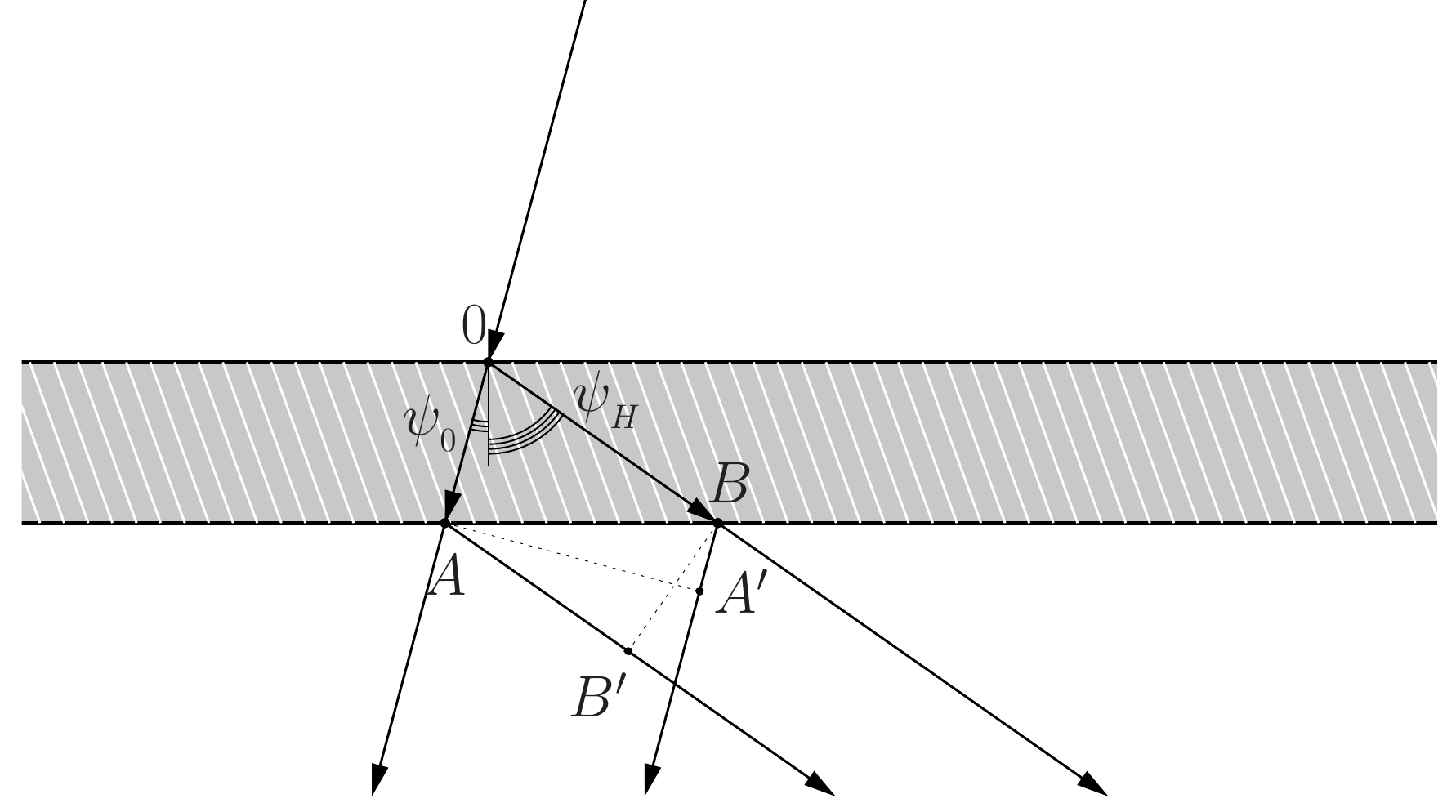}}
\end{picture}
\caption{Schematic presentation of plane wavefront paths (solid vector
  lines) in Laue-case Bragg diffraction and forward diffraction from a
  crystal plate. The total duration of forward diffraction
  $(0BA^{\prime}-0A)/c=\thicknessduration$ \eqref{thicknessduration},
  while total duration of diffraction is
  $(0AB^{\prime}-0B)/c=\thicknessduration/b$.  The lateral spread of
  forward Bragg diffraction is $AA^{\prime}=\thick \cos
  2\theta/\gammah$, and the lateral spread of Bragg diffraction is
  $BB^{\prime}=\thick \cos 2\theta/\gammao$.}
\label{fig008l}
\end{figure}

\begin{figure}[t!]
\setlength{\unitlength}{\textwidth}
\begin{picture}(1,0.27)(0,0)
\put(0.0,0.00){\includegraphics[width=0.49\textwidth]{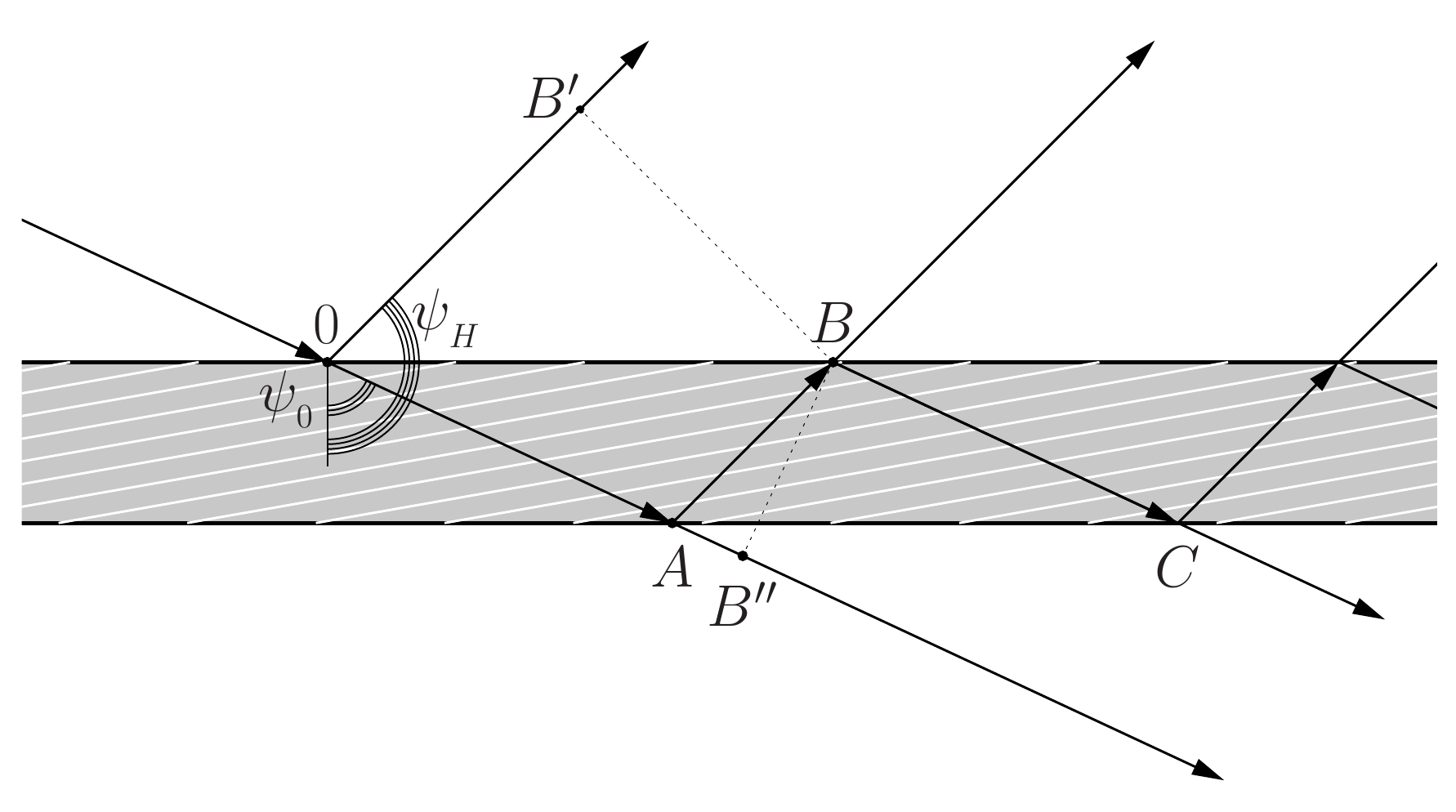}}
\end{picture}
\caption{Schematic presentation of plane wavefront paths (solid vector
  lines) with possible multiple internal reflections from the rear and
  front crystal surfaces in Bragg-case Bragg diffraction and forward
  Bragg diffraction from a crystal plate. The forward diffracted wave
  $0ABC$ originating from the front surface reflection in $B$ is
  delayed by $(0AB-0B^{\prime\prime})/c=\thicknessduration$
  \eqref{thicknessduration}, as compared to the primary forward diffracted wave
  $0B^{\prime\prime}$.  The diffracted wave $0AB$ originating from the
  rear surface reflection in $A$ is delayed by
  $(0AB-0B^{\prime})/c=\thicknessduration/|b|$ compared to the primary
  diffracted wave $0B^{\prime}$.  }
\label{fig008}
\end{figure}

The variable $\tauoo$ is equivalent to $\tauo$, while the meaning of
$\tauho$ may not be immediately evident. To gain more insight, we
rewrite $\tauho$ in an equivalent form using
\eqref{pro023}-\eqref{pro024}, \eqref{asymmetryfactor}, and Bragg's
law $\omegab=Hc/(2\sin\theta)$
\eqref{bragg}:
\begin{equation} \label{thicknessduration}
\begin{split}
\tauho\,= & \,\tauh + 2 \sin\theta\left(\frac{\vc{H}}{H}-\frac{\sin\theta}{\gammah}
\hat{\vc{z}}
 \right)\frac{\rvc}{c}+ \thicknessduration\, \deltabl\\
\thicknessduration \, = & \, \frac{2\,\thick\,\sin^2\theta}{c|\gammah |},\\
\deltabl\, = &\, \left\{ \begin{array}{ll}  
0 & \mbox{in~Bragg~geometry} \\ 
1 & \mbox{in~Laue~~geometry}.  \\
\end{array}\right.
\end{split}
\end{equation}
The parameter $\thicknessduration$ is a characteristic measure of time
in Bragg diffraction associated with the crystal thickness
$\thick$. In the Laue-case geometry, $\thicknessduration$ is equal to
the total duration of forward Bragg diffraction, which is given by the
difference in path lengths for the wave to propagate along
$0BA^{\prime}$ and $0A$, as shown schematically in
Fig.~\ref{fig008l}. The total duration of Bragg diffraction in
Laue-case geometry is determined by the difference in path lengths
$0AB^{\prime}$ and $0B$, which equals $\thicknessduration/b$ as can be
derived from schematic in Fig.~\ref{fig008l}.  Although the last term
in \eqref{thicknessduration} vanishes in the Bragg-case geometry, the
parameter $\thicknessduration$ continues to play an important role.
Unlike the Laue-case geometry, diffraction in Bragg-case geometry is
not limited in time, because multiple reflections from the front and
rear surfaces take place \cite{Wagner56,Laue60,Pinsker82,Authier} as
shown schematically in Fig.~\ref{fig008}. Accordingly, the parameter
$\thicknessduration$ is a characteristic measure of time associated
with crystal thickness in the Bragg-case geometry, where it measures
the time between multiple reflections as explained in the caption to
Fig.~\ref{fig008}.

Using Fig.~\ref{fig002}, we can express $\hat{\vc{z}}$ and $\vc{H}/H$
in \eqref{thicknessduration} in terms of the unit vector $\ucvch$ along the
diffraction optical axis, and the unit vector $\wcvch$ perpendicular
to the axis as follows: $\hat{\vc{z}}\,=\,\cos\psih\,\ucvch
\,+\,\sin\psih\,\wcvch$, and $\vc{H}/H\,=\,\sin\theta\,\ucvch
\,-\,\cos\theta\,\wcvch$. With these, the third term in
\eqref{thicknessduration} can be presented as
\begin{equation} \label{dirate}
\begin{split}
2 \sin\theta\left(\frac{\vc{H}}{H}-\frac{\sin\theta}{\gammah}\hat{\vc{z}}\right)\, = &\,  \dirate \wcvch, \\
\dirate\,=\,\frac{2\sin\theta\sin\eta}{\sin(\theta-\eta)}\,\equiv &
 -\,(1+b)  \tan\theta .
\end{split}
\end{equation}
Eq.~\eqref{thicknessduration} for $\tauho$ thus can be now written as
\begin{equation}
\tauho\,=\,\tauh + \dirate \frac{\wcvch\rvc}{c}+ \thicknessduration\,\deltabl , 
\label{pro028}
\end{equation}
where $\tauh\,=\,t-{\ucvch \rvc}/{c}$ \eqref{pro023}.

The quantity $\dirate$ in \eqref{dirate}-\eqref{pro028} is the
normalized angular dispersion rate.  It is a measure of the variation
of the propagation direction $\khvc/\ko$ of the diffracted wave
\eqref{pro027} as a function of the incident photon energy $\hbar
\ko/c$, assuming a fixed direction $\kovc/\ko$ of the incident wave
vector. Indeed, using \eqref{pro027} we obtain
$\delta(\khvc/\ko)=-\dirate \wcvch\, (\delta\ko/\ko) $, see
\cite{Shvydko-SB} for details. We note that the normalized angular
dispersion rate $\dirate$ is zero only in Bragg-case symmetric
scattering geometry with $\eta=0$. In all other cases, including the
``symmetric'' Laue geometry ($\eta=90^{\circ}$), it is nonzero.
Depending on the sign of $\eta$, $\dirate$ can take positive or
negative values in the Bragg-case geometry. In the Laue-case geometry
$\eta>\theta$ and $\dirate$ is therefore always negative.

From Eq.~\eqref{pro028} we conclude that the spatiotemporal variable
$\tauho=\tauh + \dirate {\wcvch\rvc}/{c}+ \thicknessduration$ contains
in addition to $\tauh$ an important term $\dirate {\wcvch\rvc}/{c}$
which describes a spatially lateral (perpendicular to the diffraction
axis $\ucvch$) amplitude modulation $G_{\ind{0H}}(\tauho)$ of the
diffracted radiation field \eqref{pro022}. The amplitude modulation
occurs due to interference of different spectral components
propagating in different directions, which arise from the angular
dispersion due to the additional momentum transfer $\Deltah$ -
Eq.~\eqref{pro027} and \eqref{pro031}.  The effect is very generic and
vanishes only in one case, in symmetric Bragg geometry when $\eta=0$ -
Fig~\ref{fig002}(a). We will refer to this effect as the angular
dispersive lateral spatial modulation of the diffracted wavefield
(here by angle we mean the angle of reflection rather than the angle
of incidence).

\subsubsection{Nonlinear phase}

In our preceding discussion we have focused on the phase contributions
that are linear in the frequency difference $\Omega$.  While this is a
very good approximation for the wavefields in vacuum close to the
crystal surface, the nonlinear ($\sim \Omega^2$) contributions
inherent in the additional momentum transfer $\Deltah$ give rise to
additional physics over the potentially large propagation distances
between the crystal surface and any experimental sample/detector.
Here, we briefly quantify this effect and summarize its physical
origin.

Using Eq.~\eqref{asymmetryfactor}, the definition of $\psio$ and
$\psih$ given in the caption of Fig.~\ref{fig002}, along with the
definition of $\dirate$ \eqref{dirate}, the expression for the
additional momentum transfer \eqref{pro031} can be rewritten as
\begin{equation}
  \frac{\Deltah}{\omegab/c}\, =  \,\frac{2\sin^2\theta }{\gammah}\,\frac{\Omega}{\omegab}\,- \,
  	\frac{\dirate^2 }{2\gammah}\left(\frac{\Omega}{\omegab}\right)^2+\cdots \, .\label{nlt010}
\end{equation}
Including the quadratic phase dependence in Eq.~\eqref{nlt010}, the wavefields  \eqref{pro020}-\eqref{pro022} can be presented as 
\begin{gather}
  \fieldout_{\ind{\hvar}}(\rvc ,t)\,=\,  \incamp\,\ekspon{-{\mathrm i}\omegab \taux }\,X_{\ind{0\hvar}}(\tauxo), \label{nlt020} \\
  X_{\ind{0\hvar}}(\tauxo)\,=\,\int_{-\infty}^{\infty}\frac{{\mathrm
      d}\Omega}{2\pi}\,\ekspon{-{\mathrm i}\Omega
    \tauxo}\,\refl_{\ind{0\hvar}}(\Omega)\,S(\Omega) ,
  \label{nlt022}\\
  S(\Omega)=\exp\left[-\frac{{\mathrm
        i}\dirate^2\rho}{2c\omegab\gammah}
    \Omega^2\right], \label{nlt024}
\end{gather}
where $\rho=(\rvc-\rhvc)\hat{\vc{z}}$ is the shortest distance from
the observation point to the crystal surface, so that $\rho/\gammah$
is the propagation distance along the optical axis.  We note that the
expression for the nonlinear in $\Omega$ phase factor $S(\Omega)$ in
\eqref{nlt024} is in agreement with that obtained earlier by Bushuev
in \cite{Bushuev08}, despite the different approaches used.  With the
help of the Fourier convolution theorem
we obtain for
$X_{\ind{0\hvar}}(\tauxo)$:
\begin{gather}
  X_{\ind{0\hvar}}(\tauxo)\,=\,\int_{0}^{\infty}{\mathrm
      d}\xi\,G_{\ind{0\hvar}}(\xi)\,F(\tauxo-\xi) ,	\label{nlt030}\\
  F(\tauxo-\xi) = \sqrt{\frac{2c\omegab\gammah}{{\mathrm i}\dirate^2\rho}}
  	\exp\left[ \frac{{\mathrm i} c\omegab\gammah}{2\rho}\frac{(\tauxo-\xi)^2}{\dirate^2}\right].	\label{nlt032}
\end{gather}
Here $G_{\ind{0\hvar}}$ is the crystal response given by
Eq.~\eqref{pro022}, while $F(\xi)$ is the Fourier transform of
$S(\Omega)$.  The convolution is similar in form to that associated
with paraxial evolution for the field $G_{\ind{0\hvar}}(\xi)$, with
the Fourier transform of $S$ serving as the associated Green function,
$\rho/\gammah$ the propagation distance along the optical axis, and
$\xi/D$ playing the role of the ``transverse'' coordinate.

In fact, the factor $1/D$ gives the amount the inclined intensity
front will spread in time due to natural vacuum diffraction broadening
along the transverse coordinate $\ucvch$.  In symmetric Bragg
(reflection) geometry the reflected intensity and phase fronts are
parallel, $\dirate=0$ and the time structure remains invariant, while
in all other cases $D \ne 0$ and the separated, inclined intensity
fronts will tend to smear together as the distance from the crystal
$\rho$ increases.  The maximum distance over which the linear
approximation holds and we can ignore this spreading can be estimated
as
\begin{equation}
  \rho \ll   \frac{2c\gammah}{\dirate^2 \omegab}  \left( \frac{\omegab}{\OmegaB} \right)^2, \label{nlt040}
\end{equation}
where $\OmegaB$ is the typical frequency range of Bragg diffraction,
i.e., the range in $\Omega$ over which $\refl_{\ind{0\hvar}}(\Omega)$
is appreciable.  For example, if $\OmegaB/\omegab\simeq 10^{-4}$, the
radiation wavelength $\lambda=2\pi\omegab/c\simeq 1$\AA , and
$\gammah/\dirate^2 \approx 1$, the linear approximation breaks down at
the rather small distance $\rho\approx 1$ cm from the crystal. The
smaller the bandwidth of the Bragg reflection $\OmegaB$, the larger is
the distance from the crystal over which the linear approximation
holds.

In the following we will neglect these nonlinear effects due to vacuum
diffraction upon propagation away from the crystal. If they have to be
taken into account, one should replace the response function
$G_{\ind{0\hvar}}$ \eqref{pro022} with $X_{\ind{0\hvar}}(\tauxo)$
\eqref{nlt030}-\eqref{nlt032} in the equations presented below.

\subsection{Ultra-Short  Incident  Pulse with Confined  Wavefront}
\label{pencil-beam}

In the next step, we introduce an incident x-ray pencil-beam directed
along the unit vector $\ucvco(\thetac)\equiv \ucvco $. The wavefront
of the pencil-beam is bounded in the direction $\wcvco(\thetac)\equiv
\wcvco$ perpendicular to $\ucvco$ by the transverse profile
$\Pi(\wco)$, where $\wco=\wcvco\vc{r}$.  We assume that the profile
has a characteristic width of $\sigma_{\ind{\wshift}}$ and can be
written as a Fourier transform of the angular profile:
\begin{equation}
\Pi(\wco)
=\int_{-\infty}^{\infty} \frac{{\mathrm d}\thetax}{2\pi}\, \Upsilon(\thetax)
\exp\left[-\iu  \wco  (\omegab/c)(\thetax-\thetac)\right].
\label{pro070}
\end{equation}
The characteristic angular spread $\sigma_{\ind{\theta}}$ in
$\Upsilon(\thetax)$ is related to $\sigma_{\ind{\wshift}}$ by the
uncertainty relationship $\sigma_{\ind{\theta}} \sigma_{\ind{\wshift}}
\simeq c/\omegab $. In particular, for a pencil-beam of x-rays with a
photon energy $\hbar \omega \simeq 12$~keV ($\lambda = 2 \pi c /
\omega \simeq 0.1$~nm) and a lateral spread of
$\sigma_{\ind{\wshift}}\simeq 10~\mu$m, the angular spread
$\sigma_{\ind{\theta}}\simeq 10^{-5}$~rad.

An ultra-short-in-time incident pencil-beam
$\fieldinc(\vc{r},t)=\incamp\ekspon{-\iu\omegab\tauo}\,\delta(\tauo)\,\Pi(\wco)$
can be presented as a Fourier integral over $\thetax$ of the
plane-wave components \eqref{pro010}-\eqref{pro012} propagating along
directions $\ucvco (\thetax)$ at glancing angles of incidence
$\thetax$ to the atomic planes around the central angle $\thetac$:
\begin{equation}\label{pro080}
\begin{split}
\fieldinc(\vc{r},t)& = \incamp\int_{-\infty}^{\infty}\frac{{\mathrm d}\thetax}{2\pi}\, \Upsilon (\thetax)  
\ekspon{-{\mathrm i}\omegab\tauo(\thetax)}\delta\left[\tauo(\thetax)\right]\,\\
 & \equiv\, \int_{-\infty}^{\infty} \frac{{\mathrm d}\thetax}{2\pi}\, \Upsilon (\thetax)  \int_{-\infty}^{\infty}\frac{{\mathrm d}\Omega}{2\pi}\,
  \ekspon{-{\mathrm i}{\left(\omegab+\Omega\right)\tauo(\thetax)}},\\
\tauo(\thetax)&=\,t-\frac{\ucvco(\thetax)\rovc}{c}. 
\end{split}
\end{equation}
The ultra-short-in-time pencil-beam presentation \eqref{pro080} is
valid provided the $\thetax$-dependence in the delta-function
$\delta\left[\tauo(\thetax)\right]$ can be neglected. Since
$\tauo(\thetax)\,\simeq \tauo -\frac{\wco}{c}\,(\thetax-\theta)$, this
is valid if the time delays $\tauo$ we are considering are much longer
than the inverse frequency: $\tauo \gg \sigma_{\ind{\theta}}
\sigma_{\ind{\wshift}}/c \simeq 1/\omegab$.

The spatiotemporal response $\fieldout_{\ind{\hvar}}(\rvc,t)$ of the
crystal in Bragg diffraction to the excitation by the ultra-short and
laterally bound x-ray pulse \eqref{pro080}, both for diffracted
($\hvar = H$) and forward diffracted ($\hvar=0$) components, can now
be constructed as a Fourier integral over $\thetax$ of the plane wave
solutions \eqref{pro020}-\eqref{pro022}:
\begin{equation}\label{pro040}
\begin{split}
\fieldout_{\ind{\hvar}}(\rvc ,t)\,= & \,  \incamp\,
\int_{-\infty}^{\infty} \frac{{\mathrm d}\thetax}{2\pi}\, \Upsilon (\thetax)  
\ekspon{-{\mathrm i}\omegab(\thetax)\taux(\thetax) }\,G_{\ind{0\hvar}}[\tauxo(\thetax)],  \\
\taux(\thetax)\,= & \,t-\frac{\ucvcx(\thetax)\rvc}{c},\hspace{1cm}  \tauoo(\thetax)\,=  \,\tauo(\thetax), \\
\tauho(\thetax)\,= & \,\tauh(\thetax) + \dirate(\thetax) \frac{\wcvch(\thetax)\rvc}{c}+ \thicknessduration(\thetax)\,\deltabl. 
\end{split}
\end{equation}
Here we are using again an important condition that the carrier
frequency $\omegab(\thetax)$ satisfies Bragg's law 
\begin{equation}\label{pro041}
 \omegab(\thetax)\sin\thetax\,=\,Hc/2,
\end{equation}
equivalent to \eqref{bragg}, and $\alpha=0$ condition.

Since only small $\thetax-\theta$ values are significant, we can use
$\ucvcx(\thetax)=\ucvcx+\delta\ucvcx$, with
$\delta\ucvcx=-\wcvcx\delta\thetax$. Here, the vectors $\wcvcx$ are
perpendicular to $\ucvcx$ and directed as shown in
Figs.~\ref{fig002}(a) and \ref{fig002}(b). Applying this result and Taylor expanding $\taux(\thetax)$ and $\omega(\thetax)$ from \eqref{pro040} to first order in $|\thetax-\theta| \ll 1$, we obtain
\begin{equation}\label{pro050}
\begin{split}
\taux(\thetax)\,\simeq & \,\taux + \frac{\wcx}{c}\,(\thetax-\theta),\hspace{0.7cm}  \wcx \equiv \wcvcx\rxvc,\\
\omegab(\thetax)\simeq & \,\omegabc\,\left[1-(\thetax-\theta) \cot\thetac \right], \hspace{0.5cm} \omegabc\,\equiv\, \omegab(\thetac),\\
\taux(\thetax)\omegab(\thetax)\simeq & \,\omegabc\taux+\frac{\omegabc}{c}\left(\wcx - \taux c \cot\thetac\right) (\thetax-\theta).  
\end{split}
\end{equation}

$G_{\ind{0\hvar}}(\tauxo)$ is a slowly varying function compared to
$\exp(-{\mathrm i}\omegab\taux)$.  Therefore, one can neglect
dependence of $G[\tauxo(\thetax)]$ on $\thetax$ in performing
integration over $\thetax$ in \eqref{pro040} provided the lateral
shift $\wcx$ or/and angular spread $\sigma_{\ind{\theta}}$ are not too
large, so that $\taux \gg \sigma_{\ind{\theta}} \wcx /c$.  With these
assumptions, and using again
Eqs.~\eqref{thicknessduration}-\eqref{pro028}, we arrive at the
following general expression for the spatiotemporal dependence of
Bragg diffraction from a crystal, excited with an ultra-short-in-time
pencil-beam with a lateral spatial distribution $\Pi (\wco)$:
\begin{equation}\label{pro060}
\begin{split}
\fieldout_{\ind{\hvar}}(\rvc ,t)\,= & \,  \incamp\,G_{\ind{0\hvar}}(\tauxo)\,\Pi(\wcx - \taux c \cot\thetac)\, \ekspon{-{\mathrm i}\omegabc\taux },\\ 
\taux\,= & \,t - \frac{\ucvcx\rvc}{c}, \hspace{1cm} \hvar=(0,H), \\
\tauoo\,= & \,\tauo, \hspace{1cm} \tauho\,=\,\tauh + \dirate \frac{\wcvch\rvc}{c}+ \thicknessduration\,\deltabl .
\end{split}
\end{equation}
Equations~\eqref{pro060} reveal an interesting general property: the
spatiotemporal response in Bragg diffraction ($\hvar = H$) or in Bragg
forward diffraction ($\hvar=0$) is given by a product of the
corresponding plane-wave spatiotemporal response function
$G_{\ind{0\hvar}}(\tauxo)$ \eqref{pro022} and the spatiotemporal
envelope function $\Pi(\wcx - \taux c \cot\thetac)$, whose peak shifts
along $\wcvcx$ perpendicular to the optical axis $\ucvcx$ linearly in
time. Thus, a fixed relationship \eqref{pro060} exists between the
time delay of the crystal response and the peak of the lateral
shift. In other words, the time delay is mapped onto the lateral
spatial shift.

In our previous paper \cite{LS12}, we have shown that the spatial
shift takes place in symmetric Bragg diffraction in Bragg scattering
geometry.  The solution \eqref{pro060} generalizes that result to
asymmetric diffraction both in reflection (Bragg) -
Fig.~\ref{fig002}(a), and transmission (Laue) scattering geometries -
Fig.~\ref{fig002}(b). This result can be interpreted as follows: the
incident wavefield with bounded wavefront is presented in
Eq.~\eqref{pro080} as a superposition of plane waves propagating at
different angles of incidence. At a different angle, Bragg's law is
fulfilled for different photon frequency $\omegab(\thetax)$
\eqref{pro041}.  As a result, the spatiotemporal response
$\fieldout_{\ind{\hvar}}(\rvc,t)$ of the crystal in Bragg diffraction
\eqref{pro040} is a superposition of wave fields with different
carrier frequencies $\omegab(\thetax)$ \eqref{pro041}, resulting in a
lateral spatial modulation, or, equivalently, in a lateral spatial
shift. An alternative interpretation of the derived above general
relationship between the time delay and spatial shift is discussed in
Appendix~\ref{mapping}.  We will denote this effect as the lateral
spatial modulation due to Bragg's law of dispersion, to distinguish it
from the spatial modulation due to angular dispersion discussed in
Sec.~\ref{boundless}, and {refer to $\Pi()$} in Eq.~\eqref{pro060} as
Bragg's law dispersion envelope.

We conclude: the spatiotemporal response of the crystal to the
excitation with an ultra-short and laterally bounded x-ray pulse is
accompanied by lateral spatial modulations driven by two different
mechanisms: Bragg's law of dispersion and angular dispersion.  We will
illustrate manifestation of these two mechanisms using particular
cases in Sec.~\ref{response-bragg} and Sec.~\ref{response-laue}.

Finally, using Eqs.~\eqref{pro060} the spatiotemporal dependence of
the intensity of Bragg diffraction from a crystal excited with an
ultra-short-in-time pencil-beam having a lateral spatial distribution
$\Pi (\wco)$ can be calculated using
\begin{equation}
I_{\ind{\hvar}}(\rvc ,t)\, \propto \,  |\incamp|^2\,|G_{\ind{0\hvar}}(\tauxo)|^2 \,\Pi^2(\wcx - \taux c \cot\thetac). 
\label{pro090}
\end{equation}

We have derived in this section general solutions describing the
spatiotemporal response of crystals in Bragg diffraction. In each
particular case it is important to know the appropriate plane-wave
response functions $G_{\ind{0\hvar}}(\tauxo)$.  They can be calculated
numerically in the general case, and examples are discussed in the
following Sec.~\ref{response-bragg} for the reflection (Bragg)
geometry, and in Sec.~\ref{response-laue} for the transmission (Laue)
geometry.  In some cases $G_{\ind{0\hvar}}(\tauxo)$ can be calculated
analytically, in particular, for non-absorbing crystals.  In
Secs.~\ref{response-bragg} and \ref{response-laue} we will derive
analytical expressions for the response functions of non-absorbing
crystals in the general case of asymmetric diffraction $\eta\not =0$,
in Bragg and Laue scattering geometries, respectively, and perform
analysis of the spatiotemporal crystal response using these analytical
solutions.

\section{Response  in Bragg-Case Geometry}
\label{response-bragg}  
\subsection{Diffraction and Forward Diffraction Amplitudes}

We begin this section by summarizing the well-known results of the
dynamical theory of x-ray Bragg diffraction for both the forward
diffraction $\refl_{\ind{00}}$ and diffraction $\refl_{\ind{0H}}$
amplitudes measured at the rear $(z=d)$ and the front $(z=0)$ surfaces
of a crystal, respectively:
\begin{equation}\label{rta010}
\begin{split}
\refl_{\ind{00}}\,= & \, \ekspon{{\mathrm i}\deo_{\ind{\1}} \thick}
\frac{ R_{\2}\, - R_{\1} }
{R_{\2}\, - R_{\1}\ekspon{{\mathrm i}(\deo_{\ind{\1}}-\deo_{\ind{\2}}) \thick} },\\
\refl_{\ind{0H}} \, = & \,  R_{\1} R_{\2} 
\frac{1-\ekspon{{\mathrm i}(\deo_{\ind{\1}}-\deo_{\ind{\2}}) \thick}}
{R_{\2}\, - R_{\1}\ekspon{{\mathrm i}(\deo_{\ind{\1}}-\deo_{\ind{\2}}) \thick} },
\end{split}
\end{equation}
where
\begin{equation}\label{rta020}
\begin{split}
\deo_{\ind{\nu}} \thick \,= &\, \chi_{\ind{0}}\frac{\ko\thick }{2\gammao }\, + \frac{\cnsta}{2} 
Y_{\ind{\nu}}(\dev)  , \hspace{0.5cm} R_{\nu}\, = \,\cnstd Y_{\ind{\nu}}(\dev), \\  
Y_{\ind{\nu}}(\dev)\,=&\,\left(-\dev \pm \sqrt{{\dev}^2+b/|b|}\right),\\
\dev \, = &\,\frac{\ko\extleng} {2\gammao }\,\left[b\alpha \,+\,\chi_{\ind{0}} (1-b)\right],\\
\cnsta\,=& \, {\thick}/{\extleng}, \hspace{0.6cm}  \cnstd= \sqrt{{|b|\chi_{\ind{H}}\chi_{\ind{\bar{H}}}}}/\chi_{\ind{\bar{H}}}.
\end{split}
\end{equation}
and
\begin{equation}
\extleng\,=\,\frac{\sqrt{ \gammao | \gammah | }}{\sin\theta} \extlengs, \hspace{0.5cm} \extlengs\,=\,\frac{\sin\theta }{\ko\,  |P|\,\sqrt{\chi_{\ind{H}}\chi_{\ind{\bar{H}}}}  }.
\label{extleng}
\end{equation}
Here $\chi_{\ind{\hvar}}$ $(\hvar=0,H,\bar{H})$ are Fourier
coefficients of the periodic-in-space crystal electric susceptibility
$\chi(\vc{r})$.  In general, $\chi_{\ind{\hvar}}$ are very small
complex parameters. The imaginary part $\Im\{\chi_{\ind{\hvar}}\}$ is
related to the cross-section of photo-absorption, while the real part
$\Re\{\chi_{\ind{\hvar}}\}$ is primarily related to the atomic Thomson
scattering amplitude.  In many interesting cases, e.g., for Si
crystals, $\Re\{\chi_{\ind{\hvar}}\}\gg \Im\{\chi_{\ind{\hvar}}\}$. In
certain cases, e.g., for diamond or Be crystals, one can even neglect
photo-absorption to a certain extent, and assume $\chi_{\ind{\hvar}}$
to be purely real parameters.  We make this approximation in the
analytic calculations of the response functions below, although as
shown by Kato \cite{Kato68} the resulting expressions can often be
applied to absorbing perfect crystals by letting the
$\chi_{\ind{\hvar}}$ be complex if the appropriate branches of square
roots, etc. are taken.  Typically, $\Re\{\chi_{\ind{\hvar}}\}\sim
10^{-4}-10^{-7}$ for Si and diamond crystals for 5-20~keV x-rays.

The index $\nu=1,2$, identifies two possible solutions for the
correction $\deo_{\nu}$ of the in-crystal wave vector
$\kovccrystal=\kovc+\deo\hat{\vc{z}}$ with respect to the vacuum
wavevector $\kovc$.

The diffraction $\refl_{\ind{0H}}$ and forward diffraction
$\refl_{\ind{00}}$ amplitudes in \eqref{rta010} are essentially
functions of one main parameter, the normalized deviation parameter
$\dev$ \eqref{rta020}. It contains all the information on the
magnitude of the photon frequency $\omegao$, the direction of its
momentum $\kovc$ relative to diffraction vector $\vc{H}$ and to the
internal surface normal $\vc{\hat{z}}$, the asymmetry factor $b$, and
other information pertinent to scattering geometry.

The parameter $\extleng$ \eqref{extleng} in \eqref{rta020} is an
extinction length \footnote{The definition of the extinction length
  varies by a factor of $2\pi$ in the dynamical diffraction theory
  literature. Here, we define $\extleng$ in the same way as in
  \cite{Shvydko-SB}. With such a definition, the two characteristic
  measures of time in Bragg diffraction $\thicknessduration$
  \eqref{thicknessduration} and $\timex$ \eqref{rta072} have identical
  structure.  In some other references, including
  \cite{LS12,HXRSS12,Authier}, etc., the extinction length
  $\Lambda_{\ind{H}}$ is defined to be a factor $2\pi$ larger:
  $\Lambda_{\ind{H}}=2\pi\extleng $}.  In Eq.~\eqref{extleng} we also
define the extinction length in symmetric scattering geometry
$\extlengs$, for which the asymmetry angle $\eta=0$, see
Fig.~\ref{fig002}(a). In this case $\gammao=-\gammah=\sin\theta$, and
$\extleng \rightarrow \extlengs$.  An important feature of the
symmetric version $\extlengs$ is that it is invariant for a given
Bragg reflection, being independent of the photon frequency $\omegao$
or incidence angle $\theta$ to good accuracy for crystals with small
photo-absorption; $\extlengs$ is determined solely by the diffraction
vector $\vc{H}$. This can be seen from \eqref{extleng}, by using the
fact that $\chi_{\ind{H}}\propto 1/\ko^2$ and
$\sin\theta\simeq2H/\ko$.  The extinction length $\extlengs$
determines the characteristic interaction length in Bragg diffraction
from the atomic planes with diffraction vector $\vc{H}$.  Along with
the crystal thickness $\thick$, the extinction length is another
characteristic measure of length in Bragg diffraction.

With the help of the above Eqs.~\eqref{rta020}, and using expression
\eqref{rta050} for the deviation parameter $\alpha$, the following
relationship can be established between $\dev$ and $\Omega$:
\begin{equation}\label{rta070}
  \Omega\,=\,-\sgn{b}\,\frac{\dev}{\timex} \,+\, \omegaa, 
\end{equation}
where
\begin{gather}
  \timex\,=\,\frac{2\,\extleng\, \sin^2\theta}{c\, |\gammah |}\,\equiv\,  \exttimes\, \sqrt{|b|}\sin\theta  , \label{rta072}\\
  \exttimes\,=\,2\extlengs / c. \label{timex}
\end{gather}
and
\begin{equation}
  \whh = \whhs \frac{(b-1)}{2b}, \hspace{0.5cm} \whhs = -\frac{\chi_{\ind{0}}}{2\sin^2\theta}.
\label{rta074}
\end{equation}
Here, $\timex$ is the characteristic measure of time in Bragg
diffraction associated with the extinction length.  It can be directly
compared to $\thicknessduration=\cnsta\timex$
\eqref{thicknessduration}, another characteristic measure of time in
Bragg diffraction associated with the crystal thickness. We have
introduced here also the Bragg reflection invariant time constant
$\exttimes$ \eqref{timex}, which is associated with the Bragg
reflection invariant extinction length in symmetric scattering
geometry $\extlengs$.  Typically, $\extlengs\approx 1-50~\mu$m (see
Table~\ref{tab1} in Appendix~\ref{diamondreflections}), and therefore
$\exttimes \approx 50-1000$~fs.

The parameter $\whh $ in \eqref{rta074} is a Bragg's law correction
due to refraction at the vacuum-crystal interface \cite{Shvydko-SB}.
Its magnitude $\whhs $ in symmetric diffraction ($b=-1$) is a Bragg
reflection invariant, similar to the invariance of the extinction
length $\extlengs$.  While the precise value of $\whhs$ depends on the
Bragg reflection, in most cases it is very small ($\whhs \ll 10^{-4}$,
see Table~\ref{tab1} in Appendix~\ref{diamondreflections}).

Far off the region of Bragg diffraction, where the deviation parameter
$|y|\gg 1$ or equivalently when $\alpha$ \eqref{pro017} is large, the
diffraction signal is $\refl_{\ind{0H}}(\infty)=0$ \eqref{rta010}, as
expected. On the contrary,
\begin{equation}
\refl_{\ind{00}}(\infty)=\cnstc,\hspace{0.5cm} 
\cnstc \,=\, \exp\left( {\mathrm i} \chi_{\ind{0}}\frac{\ko\thick }{2\gammao }\right) \label{rta116}
\end{equation}
has a non-zero value that represents the diffraction-free transmission
amplitude of the incident radiation with refraction and
photo-absorption accounted for by $\cnstc$ through the complex
$\chi_{\ind{0}}$.  The {\em actual} forward diffraction amplitude is
therefore obtained by subtracting off the trivial $\dev$-independent
amplitude $\cnstc$ \eqref{rta116},
\begin{equation}\label{rta090}
\tilde{\refl}_{\ind{00}}=\refl_{\ind{00}}-\cnstc,
\end{equation}
resulting in $\tilde{\refl}_{\ind{00}}(\infty)=0$.  Using the {\em actual} forward diffraction amplitude \eqref{rta090}, the forward Bragg diffraction response function \eqref{pro022} can be presented as the sum
\begin{equation}\label{rta092}
\begin{split}
G_{\ind{00}}(\tauoo)  \,= & \, \cnstc \delta(\tauoo)\,+\ \tilde{G}_{\ind{00}}(\tauoo)\\
\tilde{G}_{\ind{00}}(\tauoo)\,= & \,\int_{-\infty}^{\infty}\frac{{\mathrm
      d}\Omega}{2\pi}\,\ekspon{-{\mathrm i}\Omega
    \tauxo}\,\tilde{\refl}_{\ind{00}}(\omegab+\Omega),
\end{split}
\end{equation}
so that the response is decomposed into the prompt diffraction-free
transmission $\cnstc \delta(\tauoo)$, and the delayed {\em actual}
forward diffraction response function
$\tilde{G}_{\ind{00}}(\tauoo)$. A similar approach has been used in
\cite{KAK79} to deal with time dependence of forward resonant
scattering from M\"ossbauer nuclei.

\begin{figure*}
\setlength{\unitlength}{\textwidth}
\begin{picture}(1,0.63)(0,0)
\put(0.0,0.00){\includegraphics[width=1.0\textwidth]{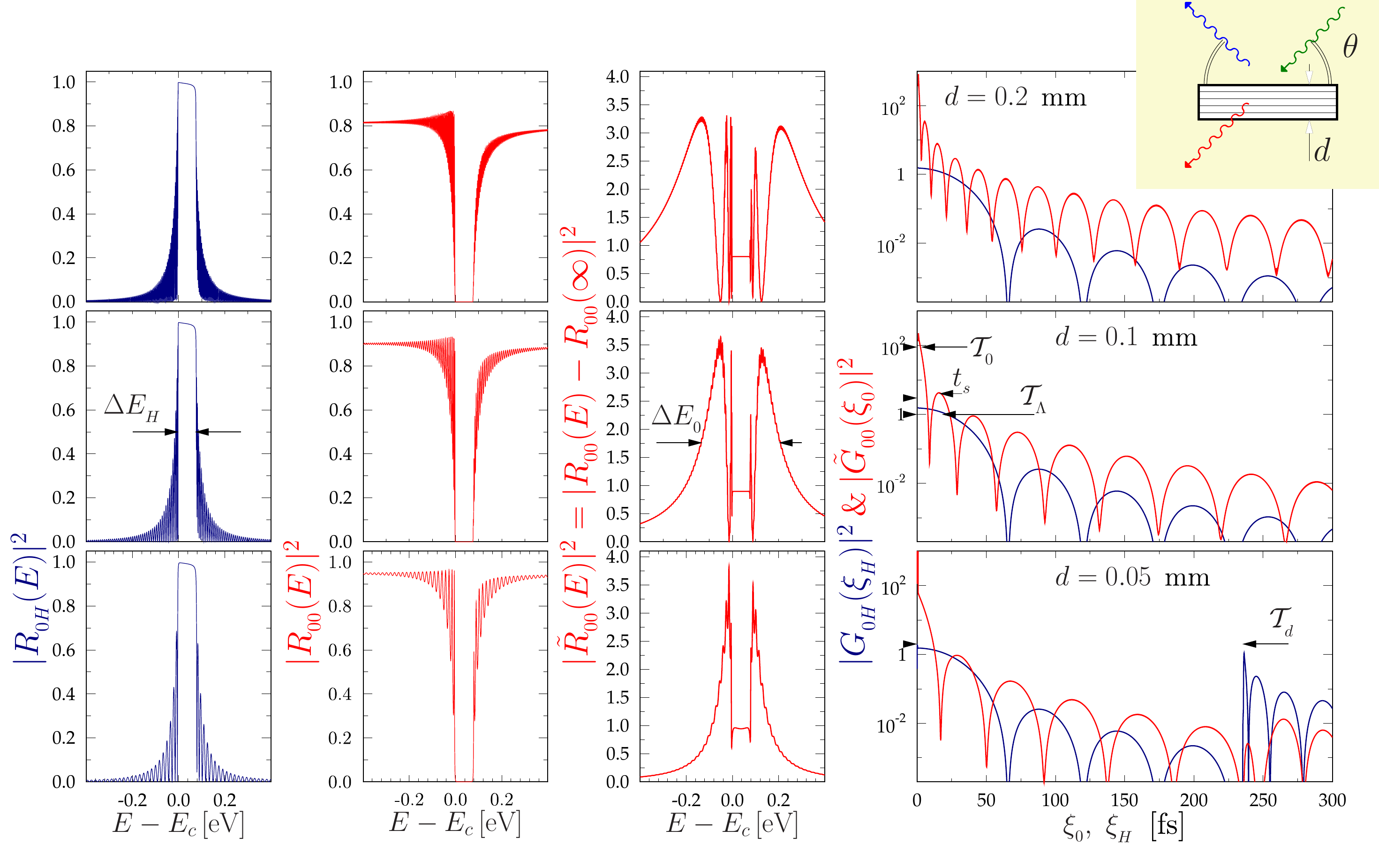}}
\end{picture}
\caption{Spectral dependences (left three columns) of
  the Bragg diffraction (BD) intensity
  $|R_{\ind{0H}}(E)|^2$, the  forward
    diffraction intensity $|R_{\ind{00}}(E)|^2$, the
  {\em actual} forward Bragg diffraction (FBD) intensity
  $|\tilde{R}_{\ind{00}}(E)|^2=|R_{00}(E)-R_{00}(\infty)|^2$, and
  the corresponding temporal intensity dependences of
  the response functions $|G_{\ind{0\hvar}}(t)|^2$
  (right column) in symmetric Bragg-case geometry.
  Numeric calculations use
  Eqs.~\eqref{rta010}-\eqref{rta074}, \eqref{rta050}, and
  Eqs.~\eqref{pro022}-\eqref{thicknessduration} for the  glancing
  angle of incidence $\theta=45^{\circ}$ to the (004) reflecting
  atomic planes in diamond, with the  asymmetry angle
  $\eta=0$. The center of the Bragg reflection region ($y=0$)
  corresponds to x-ray photon energy $E_{\mathrm c}=9.83$~keV.  The
  spectral and time dependences in diamond crystals of different
  thickness $d=0.05$~mm, $d=0.1$~mm, and $d=0.2$~mm are shown in three
  different rows from bottom to top, respectively.}
\label{fig001}
\end{figure*}

\subsection{Response Functions}

The response functions $G_{\ind{0H}}(\tauho)$ and
$\tilde{G}_{\ind{00}}(\tauoo)$ are calculated using
Eqs.~\eqref{pro022} and \eqref{rta092} with forward diffraction
$\tilde{\refl}_{\ind{00}}$ and diffraction $\refl_{\ind{0H}}$
amplitudes given by Eqs.~\eqref{rta010} and \eqref{rta090} of the
previous sections.  We use Equation \eqref{rta070} to perform the
integration over $\dev$ instead of $\Omega$ in the Fourier
integrals. In reflection (Bragg) scattering geometry $\gammah<0$, the
asymmetry ratio $b<0$, and therefore the relationship \eqref{rta070}
between $\Omega$ and $\dev$ is actually $\Omega\,=\,{\dev}/{\timex}
\,+\, \omegaa$.

The response functions $G_{\ind{0H}}(\tauho)$ and
$\tilde{G}_{\ind{00}}(\tauoo)$ can be calculated numerically in the
general case.  Figure~\ref{fig001} shows examples of such
calculations.  The left column of Fig.~\ref{fig001} shows examples of
reflectivity spectra $|\refl_{\ind{0H}}|^2$ for crystals of different
thicknesses $\thick$.  Crystals of rather large thickness are considered
$\thick \gg\extleng$. In the particular case of the $\vc{H}=(004)$
Bragg reflection in diamond crystal, the extinction length
$\extlengs=3.6~\mu$m.  The reflectivity spectra are nearly rectangular
within $|\dev | < 1$, and have an almost crystal thickness independent
form and width $\Deltaeh$. The reflectivity is almost 100\% . Such a
high reflectivity is typical for diamond crystals due to low
photo-absorption and a high Debye-Waller factor
\cite{SSC10,SSB11}. The second column in Fig.~\ref{fig001} shows
results of calculations for the forward diffraction intensity
spectra. They looks like inverse diffraction spectra, because of the
dominating contribution of the trivial transmission in the
diffraction-free region $|\dev | \gg 1$.  The third column shows
intensity spectra of the {\em actual} forward diffraction. The main
contribution is outside the region of the total Bragg reflection. The
spectral width $\Deltaeo\gg \Deltaeh$ and is crystal thickness
dependent, varying linearly with $\thick$.

The last (right) column shows the temporal dependences of the
diffraction response function intensity $|G_{\ind{0H}}(\tauho)|^2$,
and the {\em actual} forward diffraction response functions intensity
$|\tilde{G}_{\ind{00}}(\tauoo)|^2$. In agreement with the behavior of
the spectral dependences, $|G_{\ind{0H}}(\tauho)|^2$ is approximately
independent of the crystal thickness, while
$|\tilde{G}_{\ind{0H}}(\tauho)|^2$ strongly depends on the crystal
thickness $\thick$. The characteristic times of Bragg diffraction
$\timex$ and forward Bragg diffraction $\timexo$ are indicated on the
graphs.  The temporal features at $\tauxo=\thicknessduration$ for the
crystal with smallest thickness $\thick=0.05$~mm represents diffracted
and forward diffracted wavefields originating from the rear and front
surface reflections as schematically illustrated in Fig.~\ref{fig008},
c.f. also numeric calculations in \cite{SZM01,Shvydko-SB}.  In thicker
crystals these echo wavefields arrive at later times, which are
outside the presented time range. Interestingly, the characteristic
time of the Bragg diffraction response in the range $\tauho >
\thicknessduration$ changes from $\timex$ to $\timexo$.  This reflects
the fact that the contribution to this signal comes from the same
modes which contribute to forward diffraction, i.e., from the modes
propagating through the whole crystal thickness, and not from those
propagating only through the extinction length.

More insight can be obtained from analytical solutions.  The response
functions can be calculated analytically in some specific cases, e.g.,
in the approximation of a non-absorbing
($\Im\left\{\chi_{\ind{\hvar}}\right\}=0$) and thick crystal, for
which $\thick \gg\extleng$, or equivalently $\cnsta \gg 1$. In this
case, the diffraction and forward diffraction amplitudes can be
approximated by
\begin{equation}\label{rta114}
\refl_{\ind{0H}} = \cnstd\, \left\{ \begin{array}{lcl}
  -\dev + \iu\sqrt{1-{\dev}^2} & \mbox{for}
& |\dev| < 1 \\ 
 -\dev + \sgn{y}\sqrt{{\dev}^2-1} & \mbox{for} & |\dev| \geq 1 , \\
\end{array}\right.
\end{equation}
\begin{equation}\label{rta110}
\refl_{\ind{00}} = \cnstc\, \left\{ \begin{array}{lcl}
\ekspon{ -\frac{\cnsta}{2} \left( i\dev + \sqrt{1 - {\dev}^2} \right)} & \mbox{for}
& |\dev| < 1 \\ 
\ekspon{ \iu \frac{\cnsta}{2} \left( -\dev +\sgn{y}\sqrt{{\dev}^2-1} \right)} & \mbox{for} & |\dev| \geq 1 . \\
\end{array}\right.
\end{equation}
The following observations were used to obtain  Eqs.~\eqref{rta114}-\eqref{rta110}. 
In the region $\dev \geqq 1$, both $R_{\1}\ll 1$, and $R_{\1}/R_{\2} \ll 1$, while for $\dev \leqq -1$, both $R_{\2}\ll 1$, and $R_{\2}/R_{\1} \ll 1$. Neglecting these small terms, Eqs.~\eqref{rta010} transform to Eqs.~\eqref{rta114}-\eqref{rta110}.

Equation~\eqref{rta114} represents a well known result of the
dynamical theory, that Bragg diffraction from a non-absorbing, thick
crystal $\cnsta\gg 1$ takes place with total (100\%) reflectivity
$|\refl_{\ind{0H}}(\dev)|^2=1$ within the region $|\dev|< 1$, or
equivalently, using \eqref{rta070}, within the photon energy range:
\begin{equation}\label{rta200}
\Deltaeh \,=\,  2\hbar/\timex, 
\end{equation}
in agreement with results of numeric calculations shown in
Fig.~\ref{fig001}(a). Here $\Deltaeh = \hbar\Delta\Omega$.

Using Eqs.~\eqref{rta090}-\eqref{rta110}, the {\em actual} forward
diffraction amplitude can be presented in the $|y|>1$ range as
\begin{equation}\label{rta115}
\tilde{\refl}_{\ind{00}}=\cnstc\left\{\exp\left[ \iu \frac{\cnsta}{2} \left( -\dev +\sgn{y}\sqrt{{\dev}^2-1} \right) \right]-1\right\}. 
\end{equation}
Equation \eqref{rta115} is in agreement with the results of numeric
calculations shown in the third column of Fig.~\ref{fig001}. The
forward diffraction spectral width $\Deltaeo \simeq \Deltaeh
(\cnsta/2\pi) $ is a factor of $(\cnsta/2\pi)$ broader than the Bragg
diffraction spectral width, and is crystal thickness dependent.

Using the diffraction amplitude $\refl_{\ind{0H}}$ \eqref{rta114}, the forward
diffraction amplitude $\tilde{\refl}_{\ind{00}}$
\eqref{rta110}, \eqref{rta115}, and relationship
$\Omega\,=\,{\dev}/{\timex} \,+\, \omegaa$ \eqref{rta072}, we obtain
the plane-wave response functions Eq.~\eqref{pro022} in the Bragg-case
geometry (see Appendix~\ref{response-reflection} for mathematical
details):
\begin{equation}
G_{\ind{0H}}(\tauho)\,=\, \iu \frac{\cnstd }{\timex}\,
\frac{J_{1}\left({\tauho}/{\timex}\right)}{\tauho/\timex}\, \ekspon{-{\mathrm i} \omegaa  \tauho}
\label{rta124}
\end{equation}
\begin{equation}\label{rta125}
  \tilde{G}_{\ind{00}}(\tauoo) 
=  -\frac{\cnstc}{2\timexo}\,
\frac{J_{\ind{1}}\left[\sqrt{\frac{\tauoo}{\timexo} \left(1+\frac{\tauoo}{\thicknessduration}\right)}\, \right]}
{\sqrt{\frac{\tauoo}{\timexo}\left(1+\frac{\tauoo}{\thicknessduration}\right)}}\,  
\ekspon{-{\mathrm i} \omegaa  \tauoo},
\end{equation}
\begin{equation}\label{timexo}
\timexo\,=\,\timex/\cnsta\, \equiv\, \frac{2[\extlengs]^2}{c (\thick/\gammao)}.    
\end{equation}
These solutions are valid if $\tauxo < \thicknessduration $
($\hvar=0,H$), i.e., over the duration of time that is less than the
total propagation time through the crystal $\thicknessduration$. Thus,
this solution does not include possible reflections from the rear and
front crystal surfaces.

In the limit of symmetric Bragg scattering, the response function
envelopes are the same as those obtained in \cite{LS12}.  In general,
however, an asymmetric geometry changes the characteristic time
constants.

\begin{figure*}[t!]
\setlength{\unitlength}{\textwidth}
\begin{picture}(1,0.79)(0,0)
\put(0.0,-0.01){\includegraphics[width=0.99\textwidth]{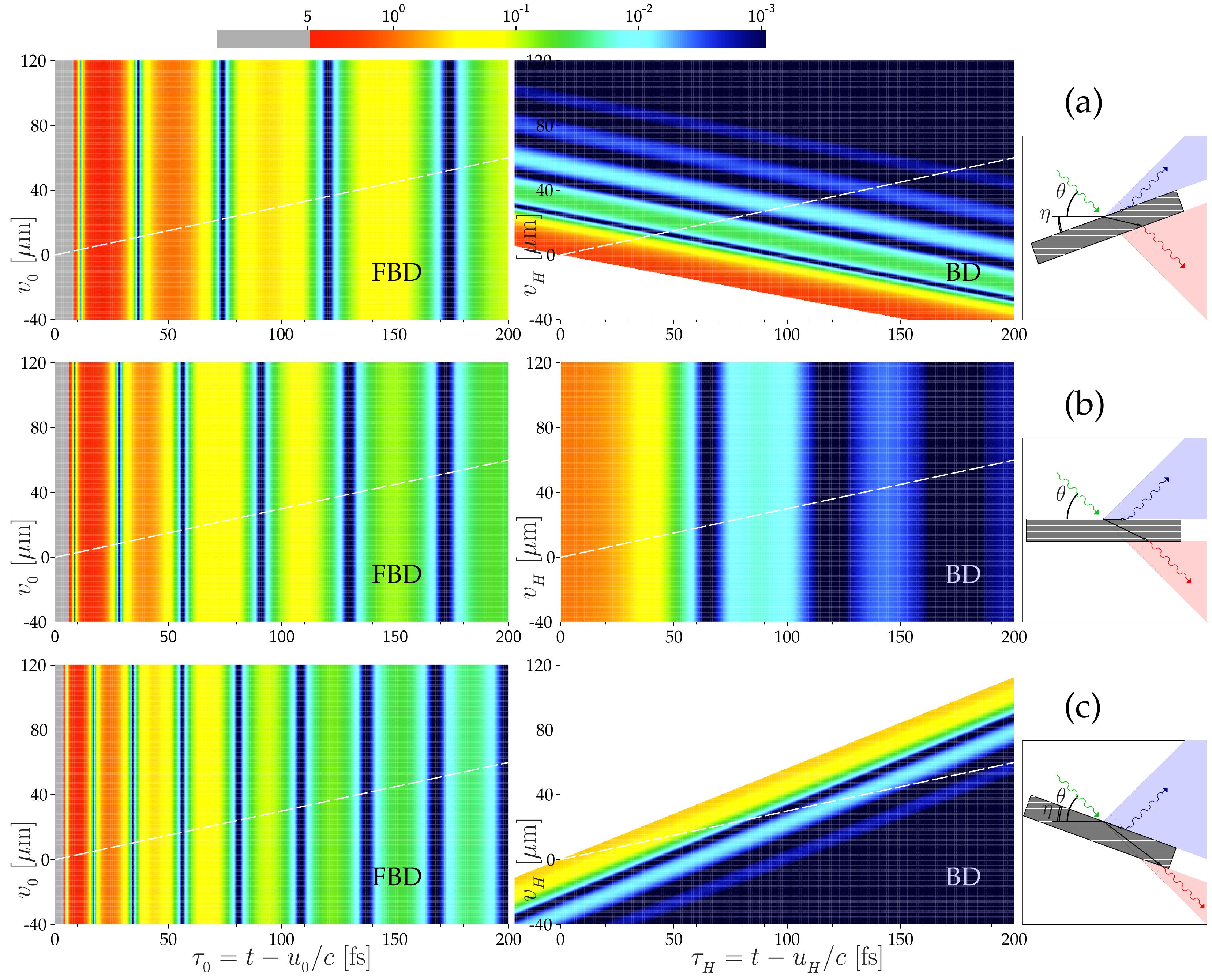}}
\end{picture}
\caption{Spatiotemporal intensity profiles of FBD and BD from a
  $100~\mu$m thick diamond crystal from the (004) Bragg reflection
  ($\extlengs=3.6~\mu$m), in the reflection (Bragg) scattering
  geometry - Fig.~\ref{fig002}(a) - with asymmetry angles
  $\eta=20^{\circ}$ (a), $\eta=0^{\circ}$ (b), $\eta=-20^{\circ}$
  (c). We plot Eq.~\eqref{pro090} using the plane-wave response
  functions \eqref{rta124}-\eqref{rta125} and a Gaussian lateral
  spatial profile of the incident x-ray beam with
  $\sigma_{\ind{\wshift}}=1000~\mu$m (i.e., a practically unbounded
  incident wavefront), were used in the calculations. Other
  parameters: $\theta=45^{\circ}$, $E=9.8$~keV, which are the same as
  those used for the calculations of the response functions shown in
  Fig.~\ref{fig001}. The intensity front tangent in Bragg diffraction
  (BD) is ${\mathrm d}\wcx/{\mathrm d}\taux = -c/\dirate$
  \eqref{dirate}. White dashed lines are traces of the Bragg's law
  dispersion envelopes $\Pi^2(\wcx - \taux c \cot\thetac)$
  \eqref{pro060}-\eqref{pro090} with a tangent ${\mathrm
    d}\wcx/{\mathrm d}\taux = c \cot\thetac$,
  c.f. Fig.~\ref{fig006ooo-10}.}
\label{fig006ooo-1000}
\end{figure*}

\begin{figure*}[t!]
\setlength{\unitlength}{\textwidth}
\begin{picture}(1,0.79)(0,0)
\put(0.0,-0.01){\includegraphics[width=0.99\textwidth]{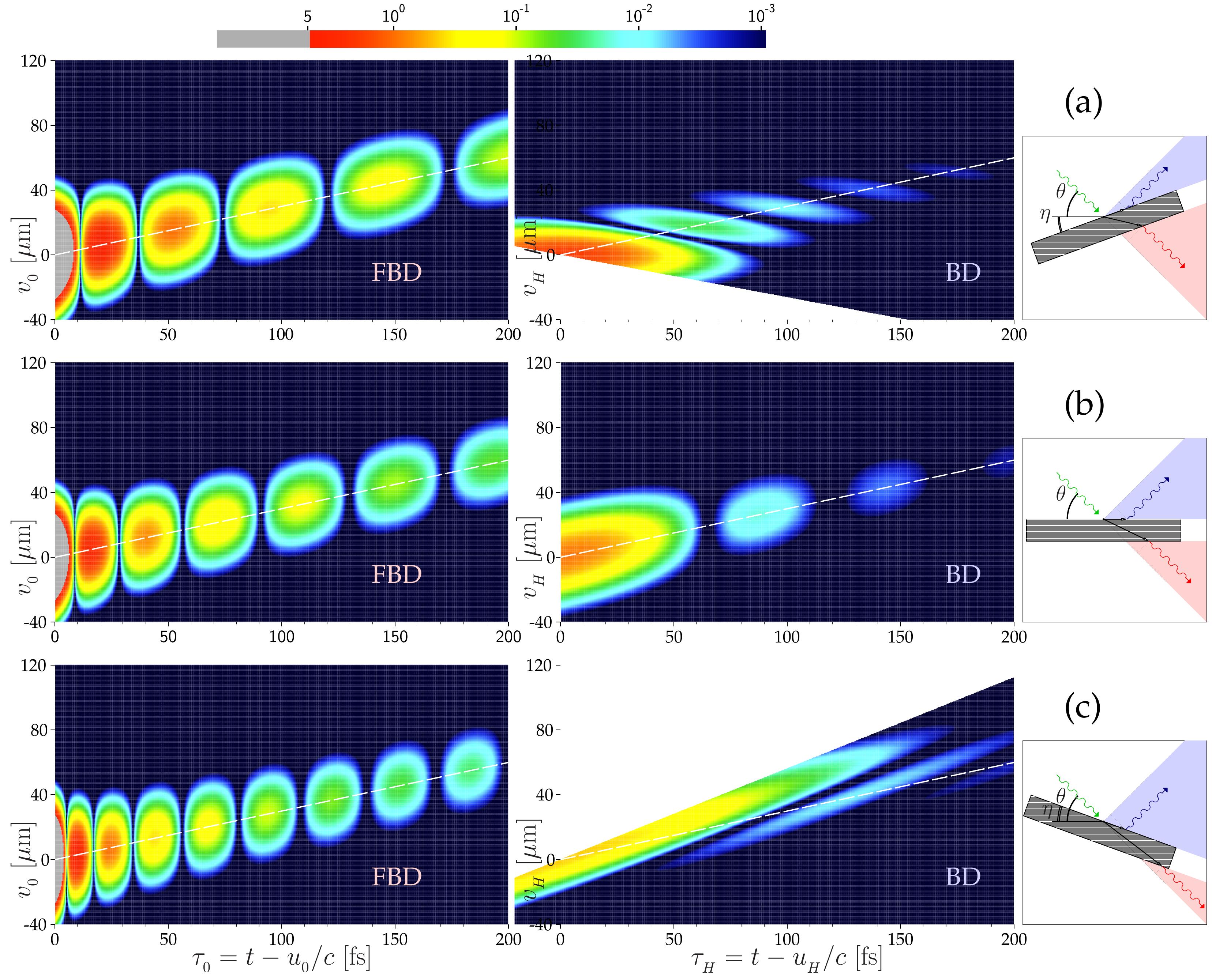}}
\end{picture}
\caption{Spatiotemporal intensity profiles of FBD and BD for similar
  parameters as that of Fig.~\ref{fig006ooo-1000}. Here, however, the
  incident x-rays have a bounded lateral spatial profile $\Pi(\wco)$
  that is assumed to be Gaussian with
  $\sigma_{\ind{\wshift}}=10~\mu$m.   Due to this, the
    spatiotemporal intensity profiles shown in
    Fig.~\ref{fig006ooo-1000} are limited now by a tight Bragg's law
    dispersion envelope $\Pi^2(\wcx - \taux c \cot\thetac)$
    \eqref{pro090} with a tangent ${\mathrm d}\wcx/{\mathrm d}\taux =
    c \cot\thetac$, shown as white dashed line.}
\label{fig006ooo-10}
\end{figure*}

According to Eq.~\eqref{rta124}, the characteristic time constant in
Bragg diffraction is $\timex$ \eqref{rta072}, which is a function of
the asymmetry factor $b$; in fact, it scales with $\sqrt{|b|}$.  By
appropriately choosing the asymmetry factor $b$, the time response can
be made faster or slower compared to the time response of Bragg
diffraction in symmetric geometry. Additionally, the uncertainty-type
relationship \eqref{rta200} associates the characteristic time
constant of diffraction with its spectral width.  As a consequence of
Eqs.~\eqref{rta200} and \eqref{rta072}, a well know result of the
dynamical theory can be reproduced: the Bragg reflection spectral
width scales with $1/\sqrt{|b|}$, $\Deltaeh =\hbar
c/(\extlengs\sqrt{|b|}\sin\theta)$. Note that both the typical energy
and time scales are predominantly determined by a single parameter,
namely, the extinction length $\extlengs$. This fact explains why
Bragg diffraction is not instantaneous, as it builds by multiple,
coherent scattering of x-rays within the extinction length.

The characteristic time constant of forward Bragg diffraction is
significantly different, since different multiple scattering processes
are involved. According to Eqs.~\eqref{rta125}-\eqref{timexo}, the characteristic time
is given by $\timexo$, which is a factor of $\cnsta=\thick/\extleng$
smaller than the characteristic constant of Bragg diffraction
$\timex$. Interestingly, $\timexo$ is practically the same as in
symmetric scattering geometry. In the general case, it is basically
defined by the Bragg reflection invariant $\extlengs$ and the
effective crystal thickness seen by incident x-rays
$\thick/\gammao$. From Eq.~\eqref{rta125}, we also calculate that the
first trailing maximum of the forward diffraction response function
appears at $t_{\ind{s}}=26\timexo$, as illustrated in
Fig.~\ref{fig001}, and its duration is $\Delta
t_{\ind{s}}=16.5\timexo$.

We note also that, if $\tauoo\ll \thicknessduration$, the expression for
the forward diffraction response function given in Eqs.~\eqref{rta125}-\eqref{timexo}
can be simplified to
\begin{equation}\label{rta125p}
  \tilde{G}_{\ind{00}}(\tauoo) 
=  -\frac{\cnstc}{2\timexo}\,
\frac{J_{\ind{1}}\left(\sqrt{{\tauoo}/{\timexo}}  \right)}
{\sqrt{{\tauoo}/{\timexo}}}\,  
\ekspon{-{\mathrm i} \omegaa  \tauoo}.\\
\end{equation}
\vspace*{-0.5cm}

\subsection{Analysis of the Spatiotemporal Response in Bragg-case Geometry}

By combining the analytical expressions for the plane-wave response
functions \eqref{rta124}-\eqref{rta125} obtained in the previous
section with the general solutions \eqref{pro060}-\eqref{pro090}, we
are now in a position to describe the spatiotemporal response of
crystals in x-ray Bragg diffraction resulting from the excitation by
an ultra-short and laterally confined x-ray pulse.

To make the analysis more instructive, we show in
Figs.~\ref{fig006ooo-1000} and \ref{fig006ooo-10} examples of 2D
($\taux,\wcx$) color plots of the spatiotemporal intensity profiles of
forward Bragg diffraction (FBD) and Bragg diffraction (BD) from a
$100~\mu$m thick diamond crystal in the (004) Bragg reflection
($\extlengs=3.6~\mu$m) with the asymmetry angle $\eta=20^{\circ}$ (a),
$\eta=0^{\circ}$ (b), or $\eta=-20^{\circ}$ (c).

Fig.~\ref{fig006ooo-1000} shows examples of calculations that apply
the above mentioned equations to an incident wavefront that is, for
all practical purposes, laterally unbounded (we assume that the
incident spatial profile has a Gaussian distribution with
$\sigma_{\ind{\wshift}}=1000~\mu$m).  In the symmetric case -
Fig.~\ref{fig006ooo-1000}(b) - the spatiotemporal profiles of both FBD
and BD are homogeneous in the lateral spatial shift $\wcx$, i.e., they
show no variation along the plane perpendicular to the appropriate
optical axis $\ucvcx$ - see Fig.~\ref{fig002}(a). The FBD response
remains independent of $\wcx$ for non-zero values of the asymmetry
angle $\eta\not =0$, as the plots demonstrate in the left columns of
Figs.~\ref{fig006ooo-1000}(a) and (c). In contrast, the BD profiles
acquire modulations along $\wch$ if $\eta\not =0$. They also produce
the impression that the wavefronts of the BD wavefields are
inclined. There are two phase factors in the expression for the
wavefield $\fieldout_{\ind{H}}(\rvc ,t)$ in Eq.~\eqref{pro060}. The
first is $\exp(-{\mathrm i}\omegabc\taux )$ defining the wavefront
perpendicular to the optical axis $\ucvch$, and another one
$\exp(-{\mathrm i} \omegaa \tauho )$ resulting from the plane-wave
response function $G_{\ind{0H}}(\tauho)$ \eqref{rta124}. Since the
second contribution is due to a small refractive correction, the
wavefront is practically not inclined.  The pronounced effect seen in
Figs.~\ref{fig006ooo-1000}(a) and (c) is actually the inclined
amplitude (intensity) front due to the amplitude modulation
perpendicular to $\ucvch$ resulting from angular dispersion, discussed
in Sec.~\ref{boundless}.  Formally, the inclination and modulation
reveal themselves through the argument $\tauho$ of
$G_{\ind{0H}}(\tauho)$ which depends both on time $t$, and, if
$\eta\not =0$, also on the space variable $\wch$ - Eq.~\eqref{pro028}.
The magnitude of the inclination to $\ucvch$ is $\dirate/c$, it scales
with the normalized angular dispersion rate $\dirate$. The inclination
of the intensity front changes sign with the sign of $\eta$.  The
tilting of the intensity profiles due to Bragg diffraction was
previously noted by Bushuev \cite{Bushuev08}.

In all cases, varying the magnitude and sign of the asymmetry angle
$\eta$ changes the time constants $\timexo$ and $\timex$, resulting in
either dilation - Fig.~\ref{fig006ooo-1000}(a) - or contraction -
Fig.~\ref{fig006ooo-1000}(c) - of the oscillating intensity structures
associated with the spatiotemporal response; this dilation or
contraction as compared to the symmetric case occurs both in the
$\wcx$ and the $\taux$ directions, as shown in
Fig.~\ref{fig006ooo-1000}(b).

In the next step, we narrow considerably the lateral spatial profile
of the incident x-ray beam. Figure~\ref{fig006ooo-10} shows examples
of calculations for incident x-rays having a Gaussian lateral spatial
profile with $\sigma_{\ind{\wshift}}=10~\mu$m, with all other
parameters being identical to those in Fig.~\ref{fig006ooo-1000}.  The
wavefield in the direction perpendicular to the optical axis $\ucvch$
is bounded by the Bragg's law dispersion envelope $\Pi^2(\wcx - \taux
c \cot\thetac)$ - Eqs.~\eqref{pro060}-\eqref{pro090}.  White dashed
lines in Figs.~\ref{fig006ooo-1000} and \ref{fig006ooo-10} are traces
of the envelope. The tangent ${\mathrm d}\wcx/{\mathrm d}\taux = c
\cot\thetac$ is independent on whether the geometry is symmetric
$\eta=0$ - Fig.~\ref{fig006ooo-10} (b), or asymmetric $\eta\not =0$ -
Fig.~\ref{fig006ooo-10} (a),(c). As has been mentioned in
Sec.~\ref{pencil-beam} this is a result of Bragg's law dispersion due
to angular spread in the incident beam caused by the bounded
wavefront.

The two effects of lateral amplitude modulation of the wavefield, due
both to angular dispersion and to Bragg's law of dispersion, can be
clearly distinguished by comparing the spatiotemporal profiles in
Fig.~\ref{fig006ooo-1000} and in Fig.~\ref{fig006ooo-10}.

\section{Response  in Laue-Case Geometry}
\label{response-laue}  

\subsection{Diffraction, Forward Diffraction Amplitudes and Response Functions}

The wavefield amplitudes in transmission (Laue) geometry are given by the following expressions \cite{Zach,Laue60,BC64,Pinsker,Pinsker82,Authier}:

\begin{equation}\label{lta010}
\begin{split}
\refl_{\ind{00}}\,=& \,\frac{1}{R_{\2}\, - R_{\1}}\,\left( R_{\2}\,\ekspon{{\mathrm i}\deo_{\ind{\1}}\thick}  - R_{\1} \ekspon{{\mathrm i}\deo_{\ind{\2}} \thick}   \right), \\
\refl_{\ind{0H}}\,= &\,  
\frac{R_{\1}R_{\2}}{R_{\2}\, - R_{\1}}\,\left( \ekspon{{\mathrm i}\deo_{\ind{\1}}\thick}  - \ekspon{{\mathrm i}\deo_{\ind{\2}} \thick}   \right).
\end{split}
\end{equation}
The notation is the same as in \eqref{rta020}, but in contrast to the reflection (Bragg) geometry, the asymmetry factor $b$ \eqref{rta020} is positive in transmission geometry. Using Eqs.~\eqref{rta020}-\eqref{rta074}, the wavefield amplitudes \eqref{lta010} can be presented as
\begin{equation}\label{lta020}
\begin{split} 
\refl_{\ind{00}}&= \, \cnstc\ \ekspon{-{\mathrm i}\cnsta\dev /2  }\,\, W(y), \\
W(y)&=\cos \left( \frac{\cnsta}{2} \sqrt{\dev^2+1} \right)+{\mathrm i}\dev\,\frac{ \sin\left(\frac{\cnsta}{2} \sqrt{\dev^2+1}\right) }{\sqrt{\dev^2+1}},
\end{split}
\end{equation}
\begin{equation}\label{lta025}
\begin{split} 
\refl_{\ind{0H}}&=  \iu \, \cnstc\ \cnstd\ \ekspon{-{\mathrm i}\cnsta\dev/2   }\,\, V(y),\\ 
V(y)&=  \frac{\sin\left(\frac{\cnsta}{2}\sqrt{\dev^2+1}\right)}{\sqrt{\dev^2+1}}.
\end{split}
\end{equation}

\begin{figure*}
\setlength{\unitlength}{\textwidth}
\begin{picture}(1,0.63)(0,0)
\put(0.0,0.00){\includegraphics[width=1.0\textwidth]{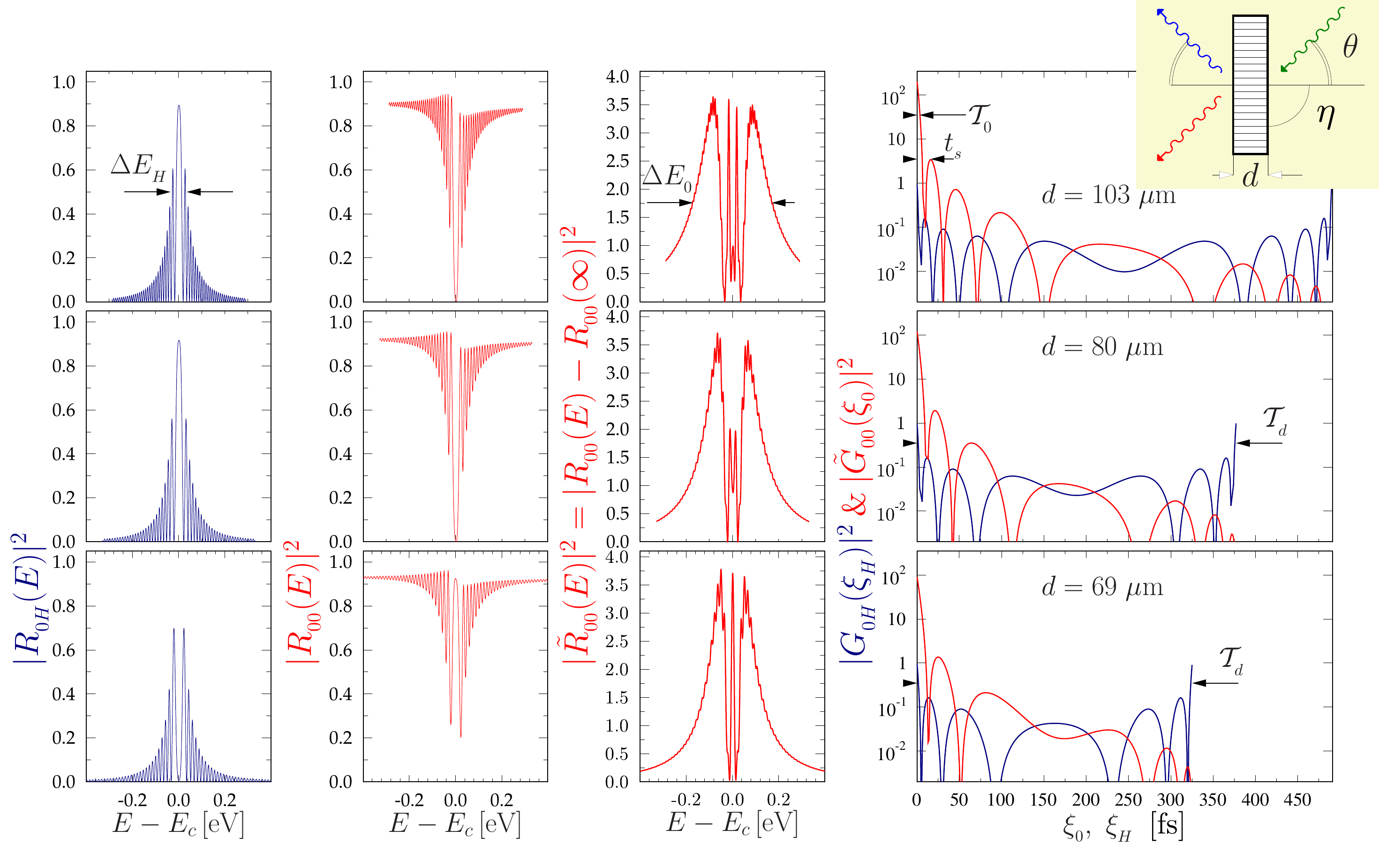}}
\end{picture}
\caption{Spectral dependences (left three columns) of the Bragg
  diffraction (BD) intensity $|R_{\ind{0H}}(E)|^2$, the forward Bragg
  diffraction intensity $|R_{\ind{00}}(E)|^2$, the {\em actual}
  forward Bragg diffraction (FBD) intensity
  $|\tilde{R}_{\ind{00}}(E)|^2=|R_{00}(E)-R_{00}(\infty)|^2$, and the
  corresponding temporal intensity dependences of the response
  functions $|G_{\ind{0\hvar}}(\tauxo)|^2$ (right column) in Laue-case
  geometry.  Numeric calculations use Eqs.~\eqref{lta010},
  \eqref{rta020}-\eqref{rta074} and \eqref{pro022} for the glancing
  angle of incidence $\theta=45^{\circ}$ to the (004) reflecting
  atomic planes in diamond, with the asymmetry angle
  $\eta=90^{\circ}$. The center of the Bragg reflection region ($y=0$)
  corresponds to the x-ray photon energy $E_{\mathrm c}=9.83$~keV.
  The spectral and time dependences in diamond crystals of different
  thickness $d=67~\mu$m, $d=80~\mu$m, and $d=104~\mu$m are shown in
  three different rows from bottom to top, respectively.}
\label{fig009}
\end{figure*}

\begin{figure*}[t!]
\setlength{\unitlength}{\textwidth}
\begin{picture}(1,1.11)(0,0)
\put(0.0,-0.01){\includegraphics[width=0.99\textwidth]{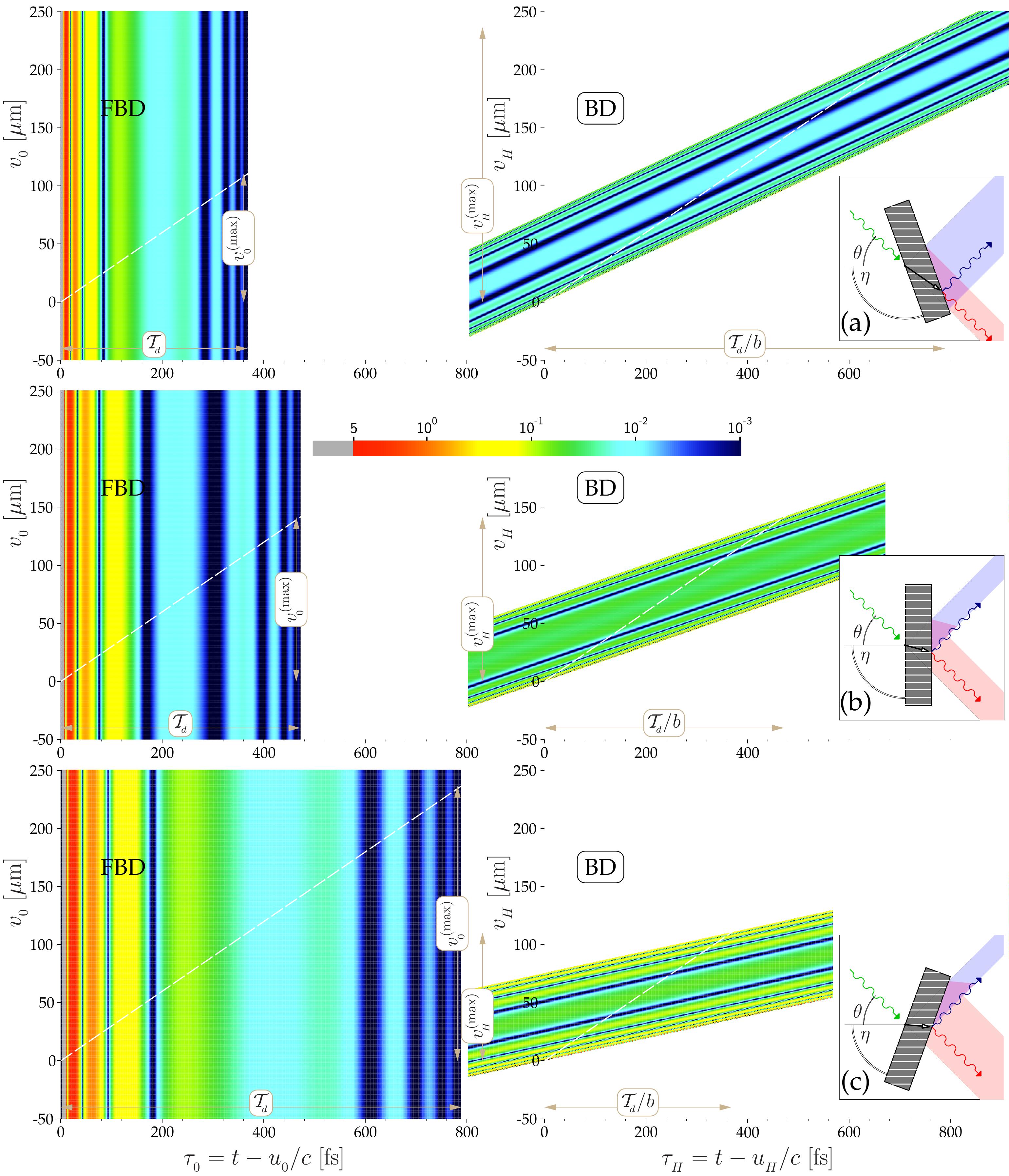}}
\end{picture}
\caption{Spatiotemporal intensity profiles of FBD and BD from a
  $100~\mu$m thick diamond crystal from the (004) Bragg reflection
  ($\extlengs=3.6~\mu$m), in the transmission (Laue) scattering
  geometry - Fig.~\ref{fig002}(b) - with asymmetry angles
  $\eta=110^{\circ}$ (a), $\eta=90^{\circ}$ (b), $\eta=70^{\circ}$
  (c). We plot Eq.~\eqref{pro090} with the plane-wave response
  functions \eqref{lta050}-\eqref{lta052} and a Gaussian lateral
  spatial profile of the incident x-ray beam with
  $\sigma_{\ind{\wshift}}=1000~\mu$m, (i.e., a practically unbounded
  incident wavefront) were used in the calculations. Other parameters:
  $\theta=45^{\circ}$, $E=9.8$~keV, which are the same as those used
  for response functions shown in Fig.~\ref{fig009}. The intensity
  front tangent in Bragg diffraction (BD) is ${\mathrm d}\wcx/{\mathrm
    d}\taux = -c/\dirate$ \eqref{dirate}.  White dashed lines are
  traces of the Bragg's law dispersion envelopes $\Pi^2(\wcx - \taux c
  \cot\thetac)$ \eqref{pro060}-\eqref{pro090} with a tangent ${\mathrm
    d}\wcx/{\mathrm d}\taux = c \cot\thetac$,
  c.f. Fig.~\ref{fig007ooo-10}.}
\label{fig007ooo-1000}
\end{figure*}

\begin{figure*}[t!]
\setlength{\unitlength}{\textwidth}
\begin{picture}(1,1.11)(0,0)
\put(0.0,-0.01){\includegraphics[width=0.99\textwidth]{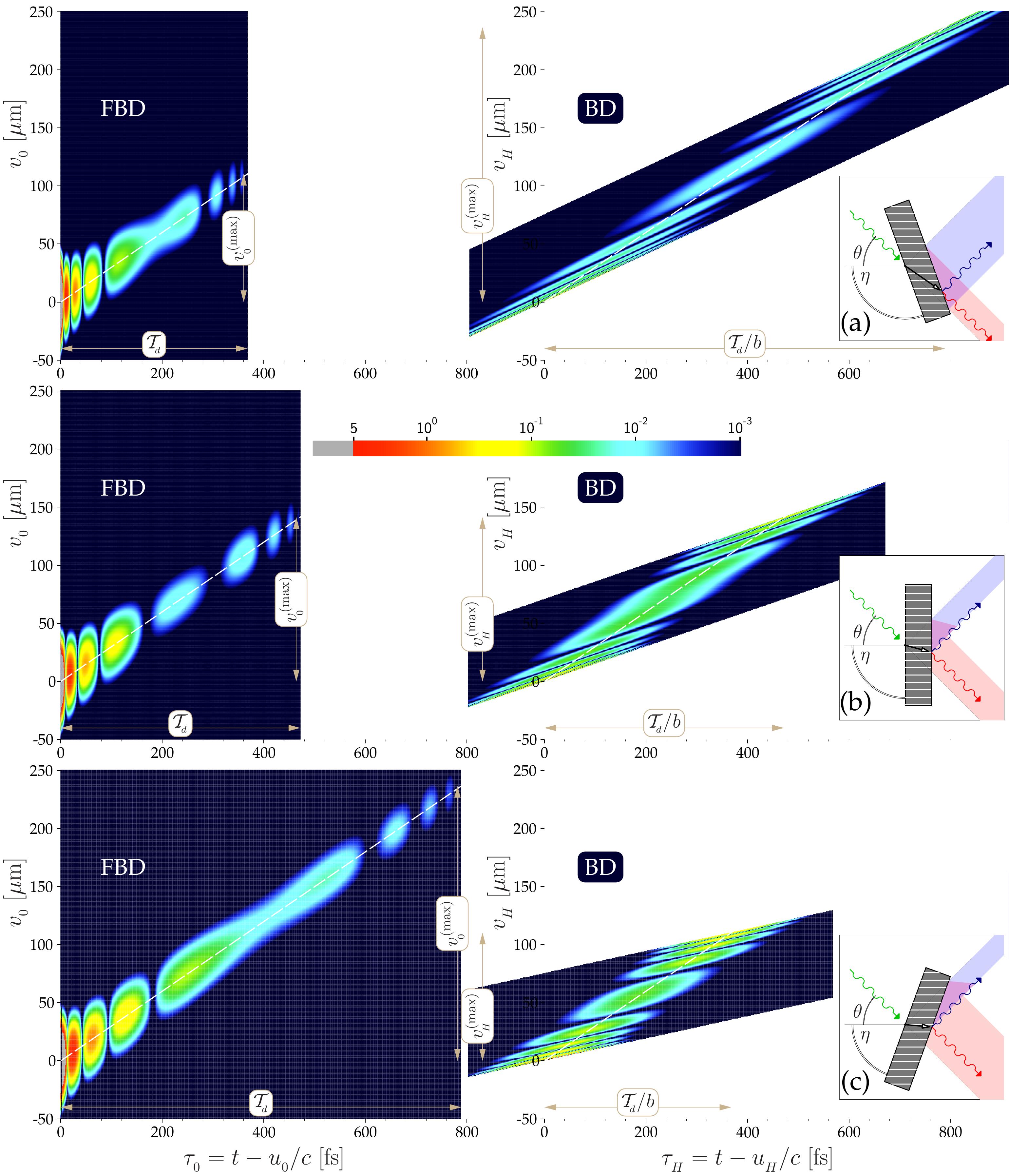}}
\end{picture}
\caption{Spatiotemporal intensity profiles of FBD and BD for similar
  parameters as that of Fig.~\ref{fig007ooo-1000}. Here, however, the
  incident x-rays have a bounded lateral spatial profile $\Pi(\wco)$
  that is assumed to be Gaussian with
  $\sigma_{\ind{\wshift}}=10~\mu$m.   Due to this, the
    spatiotemporal intensity profiles shown in
    Fig.~\ref{fig007ooo-1000} are limited now by a tight Bragg's law
    dispersion envelope $\Pi^2(\wcx - \taux c \cot\thetac)$
    \eqref{pro090} with a tangent ${\mathrm d}\wcx/{\mathrm d}\taux =
    c \cot\thetac$, shown as white dashed line.}
\label{fig007ooo-10}
\end{figure*}

Using the forward diffraction amplitude $\refl_{\ind{00}}$ given by
\eqref{lta020}, the diffraction amplitude $\refl_{\ind{0H}}$ given by
\eqref{lta025}, the relationship $\Omega=-{\dev}/{\timex} \,+\,
\omegaa$ \eqref{rta072}, and assuming zero photoabsorption, we compute
the plane-wave response functions Eq.~\eqref{pro022} for the Laue-case
which are given by (see Appendix~\ref{response-transmission} for
mathematical details):
\begin{equation}\label{lta050}
\begin{split}
  G_{\ind{00}}(\tauoo)\,=\,&\tilde{G}_{\ind{00}}(\tauoo)\,+\,
\cnstc\, \delta ( \tauoo ),\\ 
\tilde{G}_{\ind{00}}(\tauoo)=&\frac{\cnstc}{2\timexo}  \left(1-\frac{\tauoo}{\thicknessduration} \right) 
\frac{J_{\ind{1}}\left[\sqrt{\frac{\tauoo}{\timexo} \left(1-\frac{\tauoo}{\thicknessduration}\right)} \right]}{\sqrt{\frac{\tauoo}{\timexo} \left(1-\frac{\tauoo}{\thicknessduration}\right)}} \ekspon{-{\mathrm i} \omegaa  \tauoo}\\
&[0< \tauoo  < \thicknessduration ],   
\end{split}
\end{equation}
\begin{equation}\label{lta052}
\begin{split}
G_{\ind{0H}}(\tauho)=& -  \iu \, \frac{\cnstc\ \cnstd}{2 \timex }
J_{\ind{0}}\left[\sqrt{\frac{\tauho}{\timexo} \left(1-\frac{\tauho}{\thicknessduration}\right)} \right]
 \ekspon{-{\mathrm i} \omegaa  \tauho}\\
&[0< \tauho  < \thicknessduration ] .
\end{split}
\end{equation}
  Here we use characteristic time constants defined previously:
  $\thicknessduration$ in Eq.~\eqref{thicknessduration}, $\timex$ in
  Eq.~\eqref{timex}, and $\timexo$ in Eq.~\eqref{timexo}.  Both
  equations \eqref{lta050} and \eqref{lta052} can be simplified if
  $\tauho \ll \thicknessduration$:
\begin{equation}\label{lta0501}
  \tilde{G}_{\ind{00}}(\tauoo)=\frac{\cnstc}{2\timexo}  
  \frac{J_{\ind{1}}\left(\sqrt{{\tauoo}/{\timexo}}\right)}{\sqrt{{\tauoo}{\timexo} }} \ekspon{-{\mathrm i} \omegaa  \tauoo},
\end{equation}
\begin{equation}\label{lta0521}
G_{\ind{0H}}(\tauho) = -  \iu \, \frac{\cnstc\ \cnstd}{2 \timex }
J_{\ind{0}}\left(\sqrt{{\tauho}/{\timexo}}  \right)
 \ekspon{-{\mathrm i} \omegaa  \tauho}.
\end{equation}
Remarkably, for small $\tauho \ll \thicknessduration$, the forward
diffraction plane-wave response function
$\tilde{G}_{\ind{00}}(\tauoo)$ in the Laue-case geometry
\eqref{lta0521}, and its counterpart \eqref{rta125p} in the Bragg-case
geometry, are equivalent, however, with inverted signs, as a
consequence of $\sgn{b}$ in Eq.~\eqref{rta070}.

The time constant $\timexo$ \eqref{timexo} is essentially the same in
both transmission (Laue) and reflection (Bragg) geometries. It equals
the time constant $\exttimes$ in symmetric Bragg diffraction scaled by
a ratio $\extlengs/(\thick/\gammao)$ of the symmetric extinction
length $\extlengs$ to the effective crystal thickness
$(\thick/\gammao)$, i.e., the crystal thickness seen by incident
x-rays. The time constant of forward Bragg diffraction $\timexo$
\eqref{timexo} is thus general for all symmetric or asymmetric,
transmission or reflection scattering geometries. The primary
parameter controlling the forward Bragg diffraction response is the
effective crystal thickness.

\subsection{Analysis of the Spatiotemporal Response}

Expressions \eqref{lta050}-\eqref{lta052} for the plane-wave response
functions are very similar to the analogous expressions obtained by
Graeff and Malgrange in \cite{Graeff02,MG03}, with the exception that
$G_{\ind{00}}(\tauoo)$ in \eqref{lta050} contains also the prompt
$\delta$-function contribution. It originates from the spectral
components in the incident pulse with frequencies far from the Bragg
diffraction region that propagate diffraction-free through the
crystal. In Appendix~\ref{response-transmission} we also provide some
more details on the comparison with the results of
\cite{Graeff02,MG03}.

Figure~\ref{fig009} shows results of numeric calculations of the
plane-wave response function intensities
$|G_{\ind{0\hvar}}(\tauho)|^2$ and related to them spectral
dependences $|\refl_{\ind{0\hvar}}(E)|^2$ for the Laue case.  These
calculations are provided to facilitate ``visualization'' of the
analytical solutions given by Eqs.~\eqref{lta010}-\eqref{lta052}.  The
dependences shown in Fig.~\ref{fig009} are counterparts of the
analogous Bragg-case dependences shown in Fig.~\ref{fig001}. They are
calculated under the same conditions, with the single difference being
that the asymmetry angle is now $\eta=\pi/2$ (``symmetric'' Laue
case), instead of $\eta=0$ in Fig.~\ref{fig001} (symmetric Bragg
case).

In the Laue-case, the spectral range where Bragg diffraction takes
place scales with $\Deltaeh$.  While this is similar to the Bragg-case
geometry, for Laue there is no region of total reflection.  The
dominant feature of the spectral intensity dependence in
Fig.~\ref{fig009} is the intensity oscillations, which is associated
with the well-known Pendell\"osung effect of Ewald
\cite{Ewald17}. Pendell\"osung is basically related to oscillations of
the energy flow between the forward diffracted and diffracted beams in
the Laue-case geometry. The period of oscillations scales with the
extinction length $\extleng$, which is $3.6~\mu$m in this particular
case.  For the calculations presented in the middle row the crystal
thickness $\thick$ is reduced by $2\pi\extleng=22.6~\mu$m, compared to
the $\thick$ value used in the calculations presented in the top row.
The crystal thickness $\thick$ is further decreased by
$\pi\extleng=11.3~\mu$m for the calculations presented in the bottom
row.  {Varying the crystal thickness leads to} periodic in $\thick$
and in $E$ variations of the intensities that are complementary for
the diffracted and forward diffracted signals.  Interestingly, in the
spectral range $|E-E_{_{\mathrm c}}|>\Deltaeh$, the {\em actual}
forward diffraction intensity
$|\tilde{R}_{\ind{00}}(E)|^2=|R_{00}(E)-R_{00}(\infty)|^2$ has a
structure very similar to that in the Bragg-case geometry, c.f.,
Fig.~\ref{fig009}. This is in agreement with the fact that
$\tilde{G}_{\ind{00}}(\tauoo)$ for small $\tauoo$ values are identical
(modulo the inverted sign) in Bragg-case and Laue-case geometries,
c.f. Eq.~\eqref{rta125p} and Eq.~\eqref{lta0501}.

Comparison of Eq.~\eqref{lta052}, and Eq.~\eqref{rta124}, as well as
the results of numeric calculations of $|G_{\ind{0H}}(\tauho)|^2$ in
Figs.~\ref{fig009}, and \ref{fig001}, show that the characteristic
time of diffraction in Laue-case geometry is $\timexo$ \eqref{timexo},
i.e., different from the characteristic time of diffraction $\timex$
\eqref{timex} in Bragg-case geometry. This evidences that two
different characteristic length scales are involved for these two
different diffraction cases.

Eqs.~\eqref{lta050}-\eqref{lta052} and Fig.~\ref{fig009} demonstrate a
signature feature of the Laue-case plane-wave response functions
$G_{\ind{0\hvar}}(\tauxo)$. Unlike the Bragg-case analogs,
$G_{\ind{0\hvar}}(\tauxo)$ vanish outside the range
$0>\tauxo>\thicknessduration$. This effect has been reported and
discussed by Shastri et al. \cite{SZM01,SZM01-2} using numeric
calculations, and by Graeff and Malgrange using analytical solutions
in \cite{Graeff02,MG03}.

This feature, however, deserves a more detailed discussion, as it is
in fact valid only under certain conditions, but not in general. To
illustrate this, we refer to the example 2D ($\taux,\wcx$) intensity
color plots of the spatiotemporal response of forward Bragg
diffraction (FBD) and Bragg diffraction (BD) in Laue-case geometry
shown in Figs.~\ref{fig007ooo-1000} and \ref{fig007ooo-10}, which are
analogs the Bragg-case shown in Figs.~\ref{fig006ooo-1000} and
\ref{fig006ooo-10}.

Fig.~\ref{fig007ooo-1000} shows 2D plots for practically unbounded
incident wave wavefront. Unlike the Bragg case, in the Laue-case
geometry the intensity fronts of the diffracted wavefields are always
strongly inclined. This is a consequence of the non-vanishing angular
dispersion in Laue geometry, in agreement with
Eqs.~\eqref{dirate}-\eqref{pro028}.

Laue-case FBD is truly limited in time, both for observations made at
a single point or over an extended field of view: the duration is
always $\thicknessduration$ \eqref{thicknessduration}, which can be
changed by $\thick$, $\theta$, and $\eta$. The Laue-case BD is limited
in time for an observer who measures the field at a single point, and
its duration is the same as in FBD, namely,
$\thicknessduration$. However, for an observer that collects the
reflected x-rays over some region in space, the duration of BD depends
on the extent of the field of view. For an infinite field of view the
duration is infinite.

If the incident wavefront is now strongly bounded, as assumed for
calculations of the 2D plots presented in Fig.~\ref{fig007ooo-10},
then BD is limited to the region of the Bragg's law dispersion
envelope. As a result, the duration of BD for an observer with an
infinite field of view becomes limited to $\thicknessduration /b$. We
obtained this result also by ray tracing the wavefronts in
Fig.~\ref{fig008l}, which represents the limiting case of an extremely
bounded incident wavefront.

From the above examples it is clear that there is no unambiguous
answer to the question what is the duration of x-ray diffraction in
Laue-case geometry.  Depending on the conditions of the experiment, it
can be either $\thicknessduration$, or $\thicknessduration /b$, or
even arbitrarily long.  The duration of BD in the Laue-case can be
varied not only by decreasing the crystal thickness, as was suggested
in \cite{Graeff02}, but also by varying $\thicknessduration$ or
$\thicknessduration /b$ through the asymmetry angle $\eta$ and
asymmetry factor $b$, as follows from \eqref{thicknessduration}.

The lateral spread $\wcx$ of FBD and BD in Laue-case geometry appears
to be limited to $\wcx^{{\mathrm (max)}}$ if the incident beam has a
bounded wavefront. This is a well known result of the dynamical
theory, supported by many experiments, reviewed in detail, e.g., in
\cite{Authier}. Fig.~\ref{fig007ooo-10} demonstrates how the
limited-in-time crystal response correlates with the limited lateral
spread. Using these graphs one can find that the maximal lateral
spread is given by
  \begin{equation}\label{wcowchmax}
\wco^{{\mathrm (max)}}=\thick \cos 2\theta/\gammah , \hspace{1cm} \wch^{{\mathrm (max)}}=\thick \cos 2\theta/\gammao
  \end{equation}
  The same values can be obtained using ray tracing of the
  wavefronts in Fig.~\ref{fig008l}.  In agreement with
  \eqref{wcowchmax}, we find from Fig.~\ref{fig008l}
  $AA^{\prime}=\wco^{{\mathrm (max)}}=\thick \cos 2\theta/\gammah$,
  and $BB^{\prime}=\wch^{{\mathrm (max)}}=\thick \cos
  2\theta/\gammao$.

\section{Applications}
\label{applications}


\subsection{Self-Seeding of XFELs}

Understanding spatiotemporal dependencies in Bragg diffraction of
x-rays has immediate practical implications, in particular for
self-seeding of x-ray free-electron lasers (XFELs).  The self-seeding
scheme uses an upstream XFEL to generate an intense x-ray pulse via
self-amplified spontaneous emission (SASE).  The relatively
broad-bandwidth SASE pulse is then put through an x-ray monochromator
to generate a monochromatic seed for the downstream XFEL undulators,
which in turn amplifies the narrow bandwidth seed to produce fully
coherent x-rays \cite{FSS97,SSSY}.  However, traditional two- or
four-bounce monochromators induce a large delay ($>10$~ps) of the
x-rays, which in turn requires an impractically long ($\sim 40$ m)
electron beam transfer line.

A very clever, readily realizable idea of a ``wake'' monochromator
that produces a monochromatic x-ray seed at an optimal $\approx 20$~fs
delay has been proposed by Geloni et al. \cite{GKS10,GKS11}, and
recently realized at the LCLS XFEL by an international team lead by
Emma \cite{HXRSS12}.

In the original proposal \cite{GKS10,GKS11} the authors applied the
equations of the dynamical theory of x-ray diffraction in crystals to
calculate numerically the time dependence and strength of the
monochromatic seed propagating in forward direction. The action of the
monochromator crystal in the Bragg-transmission geometry was
interpreted in terms of a Bragg diffraction (BD) band-stop filter.
The underlying physics is actually related to forward Bragg
diffraction (FBD). We have discussed in detail its properties in the
symmetric Bragg-case geometry relevant for self-seeding in
\cite{LS12}. We showed that, first, the characteristic time for FBD is
$\timexo$ \eqref{timexo}, substantially different (shorter) than the
characteristic BD time $\timex$ \eqref{timex}, and therefore the
crystal in FBD generates a seed with a broader spectrum than a BD
band-stop filter would do.  Second, it was shown that the intensity of
the monochromatic seed is $ \propto 1/\timexo^2$,
c.f. Eq.~\eqref{rta125p}, which can therefore be enhanced by varying
parameters composing $\timexo$.  Similarly, its time delay
$t_{\ind{s}}=26\timexo$ (see Fig.~\ref{fig001}), and its duration
$\Delta t_{\ind{s}}=16.5\timexo$ can be tailored by changing
$\timexo$, which can be done practically by adjusting the extinction
length $\extlengs$ (for example, by choosing another reflection or
asymmetry parameter), or by changing the crystal thickness.  A
limitation of this scheme has been also identified in \cite{LS12}. It
is due to the lateral shift of the FBD signal. This is a very generic
effect, caused by the Bragg's law dispersion, as discussed in
Sec.~\ref{pencil-beam} of the present paper.

The theory developed in the present paper allows us to diversify the
variety of possible forward diffraction self-seeding monochromator
schemes. First of all, forward Bragg diffraction in Laue-case geometry
is a competitive approach. The possibility of applying FBD in
Laue-case geometry for self-seeding becomes immediately apparent from
the derived equivalence of the forward diffraction plane-wave response
functions $|\tilde{G}_{\ind{00}}(\tauoo)|^2$ in the Laue-case geometry
\eqref{lta0521}, and of its counterpart \eqref{rta125p} in the
Bragg-case geometry. The equivalence holds for small $\tauho \ll
\thicknessduration$, which is the range most appropriate for
self-seeding of femtosecond long XFEL pulses. As has been established
in this paper, the time constant of forward Bragg diffraction
$\timexo$ \eqref{timexo} is common for all symmetric or asymmetric,
transmission or reflection scattering geometries, and is the only
parameter which defines the strength, delay, and duration of FBD and
therefore of the monochromatic seed. These properties advance FBD both
in Bragg and Laue-case geometries, including asymmetric ones, to a
universal approach for the generation of monochromatic, delayed seeds
for self-seeded XFELs.  The physics is controlled by the parameters
which compose $\timexo$ \eqref{timexo}: the magnitude of the effective
crystal thickness $\thick/\gammao$, and the extinction length
$\extlengs$ in the symmetric Bragg reflection. Table~\ref{tab1} in
Appendix~\ref{diamondreflections} provides some useful data for Bragg
reflections in diamond, which can be used to select the Bragg
reflection most appropriate for the desired application. Similar data
for silicon and Al$_2$O$_3$ crystals can be found in
\cite{Shvydko-SB}.

There is no universal answer to the question: which geometry is
better, Bragg or Laue?  We investigate certain aspects of this
question below, where for simplicity we have restricted our analysis
to the symmetric diffraction geometries, defined by $\eta=0$ in the
Bragg-case - Fig.~\ref{fig002}(a), and by $\eta=\pi/2$ in the
Laue-case - Fig.~\ref{fig002}(b).

If the experimenter highly values operating the self-seeding
monochromator over as large a spectral tuning range as is possible,
than the Laue-case geometry may be a better choice.  The strongest
variation of the photon energy $E$ with the glancing angle of
incidence to the reflecting atomic planes $\theta$ takes place for
small $\theta\lesssim \pi/6$, i.e., in the linear range of Bragg's law
$E\sin\theta=\ebr$ \eqref{bragg}. We write here Bragg's law in terms
of photon energy $E$ and Bragg energy $\ebr\,=\,Hc\hbar/2$, the
smallest photon energy for which Bragg's law can be fulfilled (at
$\theta=\pi/2$).  In the symmetric Laue-case geometry the effective
thickness ${\thick}/{\gammao}={\thick}/\cos\theta$ does not vary much
if $\theta$ is small, unlike the Bragg-case in which
${\thick}/{\gammao}={\thick}/\sin\theta$. Therefore, a large variation
in $E$ is accompanied in the Laue-case geometry with a small variation
in ${\thick}/{\gammao}$ and therefore in $\timexo$, resulting in a
rather stable seed power and time delay of the seed over a large range
of photon energies. In the Bragg-case geometry this is not the case.
In addition, the Laue-case geometry at small $\theta$ allows for using
thicker crystals for the same $\timexo$ as compared to the similar
situation in the Bragg-case. This may represent a technical advantage
since the fabrication of thin crystals is typically more challenging.

Using small $\theta$ angles, however, also has its disadvantages.  The
lateral spatial shift, given by the Bragg's law dispersion envelope
$\Pi^2(\wcx - \taux c \cot\thetac)$ - Eq.~\eqref{pro090}, is $\wcx =
\taux c \cot\thetac$, i.e., proportional to $\cot\theta$, and is
maximal in the range of small $\theta$. This may not be significant
for very short x-ray pulses that can use short delay times
$t_{\ind{s}}$. However, if one wants to seed long XFEL pulses $\simeq
50-100$ fs, then the Bragg-case scattering geometry close to
backscattering $\theta\rightarrow\pi/2$ would be a more advantageous
option, albeit at a decrease in the spectral tuning range.

\subsection{Ultra-fast Time Measurements by Mapping Time on Space }


Angular dispersion in asymmetric Bragg diffraction results in an
inclined intensity front of the diffracted wavefields. This effect is
illustrated in Figs.~\ref{fig006ooo-1000} and \ref{fig007ooo-1000}, in
Bragg-case and Laue-case geometries, respectively.  Inclination of the
intensity front in asymmetric x-ray diffraction geometry was, to our
knowledge, first explicitly derived using ray tracing in
\cite{ZHZB99}, where it was proposed to be used for x-ray pulse
compression.

\begin{figure}[t!]
\setlength{\unitlength}{\textwidth}
\begin{picture}(1,0.27)(0,0)
\put(0.0,0.00){\includegraphics[width=0.49\textwidth]{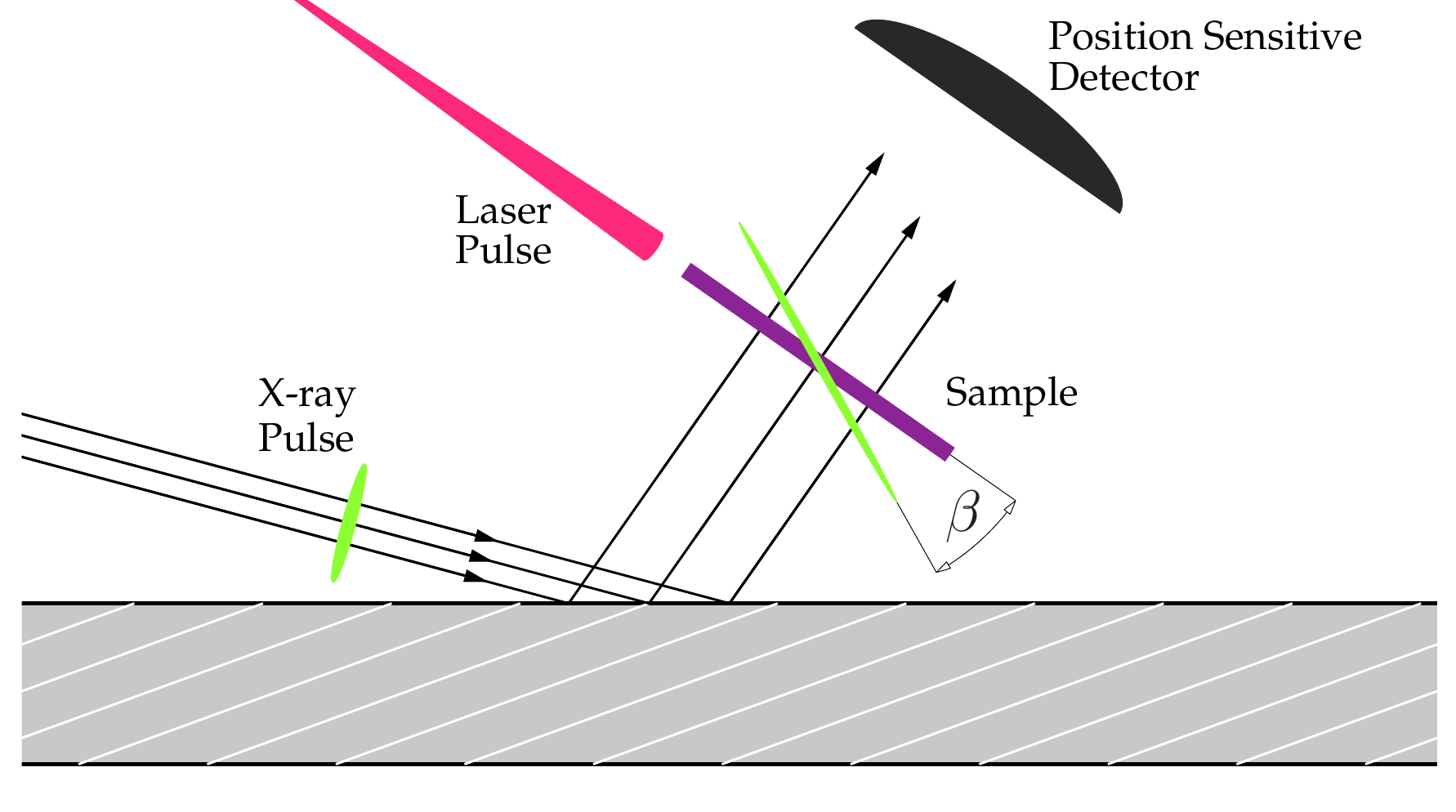}}
\end{picture}
\caption{Schematic of ultra-fast time measurements by mapping time
  delay on space in asymmetric x-ray Bragg diffraction from a
  crystal. The intensity front is rotated by an angle $\beta$ upon
  asymmetric Bragg reflection of x-rays. Here $\tan\beta=\dirate $ is
  determined by normalized dispersion rate $\dirate $ \eqref{dirate}.}
\label{fig010}
\end{figure}

Here, we suggest using the effect of intensity front inclination for
time measurements of ultra-fast processes. The schematic drawing in
Fig.~\ref{fig010} explains the idea.  The inclined intensity front
allows mapping time onto space, as different parts of the inclined
front traverse and thus probe the sample at different times.

We assume that the process under study is triggered homogeneously over
its extent by, for example, an external laser.  If we then measure the
sample with an inclined x-ray intensity front from an asymmetric
crystal, than different transverse positions will be probed at
different times, so that time dynamics can be extracted with a
spatially resolved x-ray detector. We assume that the x-ray wavefront
is sufficiently broad so that we may ignore any small additional
spatial shifts that take place due to Bragg's law dispersion.

To estimate the achievable time resolution, we first neglect the
finite duration of the incident x-ray pulse, the finite duration of
Bragg diffraction, and the sample thickness.  In this case, the
resolution of the time measurements $\Delta t=(\Delta
x/c)/\tan\beta=\Delta x/(c\dirate)$ is determined by the spatial
resolution of the detector $\Delta x$ and normalized dispersion rate
$\dirate $ \eqref{dirate}. With $\Delta x\simeq 1-10~\mu$m, and
$\dirate\simeq 1$, an estimate for the time resolution is $\Delta t
\simeq 3-30$~fs. The duration of the incident x-ray pulse, the
duration of Bragg diffraction, and the sample thickness will increase
this number. The duration of the Bragg diffraction is $\timex$ in
Bragg-case or $\thicknessduration$ in Laue-case geometry. By an
appropriate choice of $\extlengs$ and asymmetry factor $b$, the
characteristic time Bragg diffraction \eqref{timex} can be made,
however, as small as $\timex\simeq 1$~fs, i.e. smaller than the
$\Delta t$.  Tilting the sample to the x-rays propagation direction
(without tilting the intensity front) may result in a similar
effect. However, tilting the intensity front in many cases may be
advantageous, as it is decoupled from the propagation direction in the
sample, which may be an important parameter of experiments, of
diffraction experiments in particular.

\section{Conclusions}

The spatiotemporal response of crystals in x-ray Bragg diffraction
resulting from excitation by an ultra-short, laterally confined x-ray
pulse was studied theoretically. The theory developed in the paper
presents an extension of the analysis in symmetric reflection geometry
\cite{LS12} to a generic case, which includes Bragg diffraction both
in reflection (Bragg) and transmission (Laue) asymmetric scattering
geometries.

The spatiotemporal response is presented as a product of a
crystal-intrinsic plane wave spatiotemporal response function and an
envelope function defined by the crystal-independent transverse
profile of the incident beam and the scattering geometry.  The
diffracted wavefields exhibit amplitude (or intensity) modulation
perpendicular to the propagation direction due to two effects: angular
dispersion and dispersion due to Bragg's law. Angular dispersion
results in the inclination of the intensity front of Bragg diffraction
in asymmetric geometries.  Bragg's law dispersion produces a lateral
spatial shift $\wco$ of photons emerging from the crystal with respect
to the incident x-ray pulse that increases linearly with time delay
$\tauo$. A simple general relationship $c\tauo=\wco\tan\theta$ holds
in all diffraction geometries. This effect can be interpreted also in
terms of the energy flow of the wavefields in the crystal.

The spatiotemporal plane-wave response functions in Bragg diffraction
can be expressed in terms of three characteristic space and related to
them three time parameters: (i) crystal thickness $\thick$ and
$\thicknessduration$ \eqref{thicknessduration}, (ii) extinction length
$\extleng$ \eqref{extleng} and $\timex$ \eqref{timex}, (iii)
rescattering length $\extleng^2\thick$ and $\timexo$
\eqref{timexo}. The glancing angle of incidence $\theta$ and the
asymmetry angle $\eta$ also enter the three time parameters, and,
therefore, are factors that change the spatiotemporal response scale.

We address some practical applications of the developed theory.  We
show that forward Bragg diffraction (FBD) of x-rays in Laue-case
geometry can be used for self-seeding of hard x-ray free electron
lasers, along with FBD in the Bragg-case geometry.  Laue-case FBD is
advantageous if a large spectral tuning range is required. We discuss
also a possibility of using asymmetric diffraction for ultra-fast time
measurements with femtosecond resolution.

\begin{acknowledgements} The authors would like to thank Sasha
  Zholents for reading selected sections of the manuscript and
  valuable suggestions.  This work is supported by the U.S. Department
  of Energy, Basic Energy Sciences, Office of Science, under contract
  DE-AC02-06CH11357.

\end{acknowledgements}

\appendix
\section{Response Function in Reflection (Bragg) Geometry}
\label{response-reflection}  

To compute the temporal response of the forward diffracted wave, we must evaluate
\begin{widetext}
\begin{equation}
\tilde{G}_{\ind{00}}(\tauoo)\,=\, -\frac{\cnstc}{\timex}
\ekspon{-{\mathrm i} \omegaa \tauoo}
\int_{-\infty}^{\infty}\frac{{\mathrm d}\dev}{2\pi}\,
\ekspon{-{\mathrm i}\left(\tauoo/\timex \right) \dev  }\,
\tilde{R}_{\ind{00}}(y)
\label{rta130}
\end{equation}
where the $\tilde{R}_{\ind{00}}(y)$ is given by subtracting $\cnstc$
from \eqref{rta110}.  Since causality requires
$\tilde{G}_{\ind{00}}(\tauoo<0) = 0$, we have found that the most
convenient way to treat this particular problem is as an inverse
Laplace transform.  In the table of inverse Laplace transforms given
by Erd\'elyi, Magnus, Oberhettinger, and Tricomi \cite{EMOT54}, we
find that
\begin{equation}
  \int_0^\infty \! dt \; e^{-p t} f(t) = 1 - e^{-b(\sqrt{p^2 + a^2} - p)}	\label{eqn:Laplace}
	\;\;\; \Rightarrow \;\;\; f(t) = \frac{ab}{\sqrt{t(t + 2b)}} J_1\!\left[ a\sqrt{t(t + 2b)} \right].
\end{equation}
Now, we make the replacements: $t = \tauoo$, $p = -{\mathrm i}y/\timex$, $a = 1/\timex$, and $b = \cnsta\timex/2$; then, \eqref{eqn:Laplace} is proportional to the approximate Bragg transmission.  Thus, we have that
\begin{equation}
  \int_0^\infty \! d\tauoo \; e^{{\mathrm i}y\tauoo/\timex} f(\tauoo) = 
  	e^{i\cnsta/2 [\pm\sqrt{y^2 - 1} - y]} - 1 \;\; \Rightarrow \;\; 
	f(\tauoo) = -\frac{\cnsta}{2} \frac{J_1\left[\sqrt{\tauoo(\cnsta\timex + \tauoo)}/\timex\right]}
	{\sqrt{\tauoo(\cnsta\timex + \tauoo)}}. 	\label{Bragg:Append}
\end{equation}
In terms of the forward Bragg diffraction amplitude, \eqref{Bragg:Append} implies that
\begin{equation}
\tilde{G}_{\ind{00}}(\tauoo)\,=\, -\frac{\cnstc}{\timex}
\ekspon{-{\mathrm i} \omegaa \tauoo}
\int_{-\infty}^{\infty}\frac{{\mathrm d}\dev}{2\pi}\,
\ekspon{-{\mathrm i}\left(\tauoo/\timex \right) \dev  }\,
\tilde{R}_{\ind{00}}(y) = -\frac{\cnstc}{2\timexo}\,
\frac{J_{\ind{1}}\left[\sqrt{\frac{\tauoo}{\timexo} \left(1+\frac{\tauoo}{\thicknessduration}\right)}\, \right]}
{\sqrt{\frac{\tauoo}{\timexo}\left(1+\frac{\tauoo}{\thicknessduration}\right)}}\,  
\ekspon{-{\mathrm i} \omegaa  \tauoo},
\end{equation}
with $\timexo =\timex/\cnsta\, \equiv\, 2\gammao[\extlengs]^2/(c \thick)$. On the other hand, the reflected wave is given by
\begin{equation}
  G_{\ind{0H}}(\tauho)\,=\, \frac{\cnstd }{\timex}\,
  \ekspon{-{\mathrm i} \omegaa  \tauho}\,
  \int_{-\infty}^{\infty}\frac{{\mathrm d}\dev}{2\pi}\,
  \ekspon{-{\mathrm i}\left(\tauho/\timex\right) \dev  }\,
R_{\ind{0H}}(y),
\label{rta120}
\end{equation}
and the integrals are simple enough for {\sc Mathematica} to do; we find that
\begin{equation}
\begin{split}
  &\int_1^\infty \frac{{\mathrm d}\dev}{2\pi}\,
  \ekspon{-{\mathrm i}\left(\tauho/\timex\right) \dev  } \left(-\dev + \sqrt{\dev^2-1}\right) +   
  	\int_{-\infty}^{-1} \frac{{\mathrm d}\dev}{2\pi}\, \ekspon{-{\mathrm i}\left(\tauho/\timex\right) \dev  } 
	\left(-\dev - \sqrt{\dev^2-1}\right) \\
  &= 2{\mathrm i}\int_1^\infty \frac{{\mathrm d}\dev}{2\pi}\, \sin\left(\tauho \dev/\timex\right) 
  	\left(\dev - \sqrt{\dev^2-1}\right) = {\mathrm i} \frac{J_1(\tauho/\timex)}{2\tauho/\timex}
	\text{sgn}(\tauho/\timex) +
	{\mathrm i} \frac{(\tauho/\timex)\cos(\tauho/\timex) - \sin(\tauho/\timex)}{\pi(\tauho/\timex)^2}
\end{split}
\end{equation}
and
\begin{equation}
  \int_{-1}^{1} \frac{{\mathrm d}\dev}{2\pi}\, \left[{\mathrm i}\dev \sin(\tauho \dev/\timex) + 
  	{\mathrm i}\sqrt{1 - \dev^2}\cos(\tauho \dev/\timex)  \right] =
	{\mathrm i}\frac{J_1(\tauho/\timex)}{2\tauho/\timex} - 
	{\mathrm i} \frac{(\tauho/\timex)\cos(\tauho/\timex) - \sin(\tauho/\timex)}{\pi(\tauho/\timex)^2}.
\end{equation}
Adding these two, we obtain
\begin{equation}
  G_{\ind{0H}}(\tauho)\,=\, \frac{\cnstd }{\timex}\,
  \ekspon{-{\mathrm i} \omegaa  \tauho}\,
  \int_{-\infty}^{\infty}\frac{{\mathrm d}\dev}{2\pi}\,
  \ekspon{-{\mathrm i}\left(\tauho/\timex\right) \dev  }\,
R_{\ind{0H}}(y) = \iu \frac{\cnstd }{\timex}\,
\ekspon{-{\mathrm i} \omegaa  \tauho}\,
\frac{J_{1}\left(\tauho/\timex\right)}{\tauho/\timex} \Theta(\tauho/\timex)
\label{rta124b}
\end{equation}

\end{widetext}

\section{Response Function in Transmission (Laue) Geometry}
\label{response-transmission}  

To calculate response functions in Laue-case diffraction geometry we
use Eqs.~\eqref{pro022}, \eqref{lta020}, \eqref{lta025}, and
Eq.~\eqref{rta070} in the form $ \Omega=-{\dev}/{\timex} + \omegaa$,
as $b>1$ in Laue-case geometry. As a result we obtain
\begin{equation}\label{lta030}
\begin{split}
G_{\ind{00}}(\tauoo)=&  -\frac{\cnstc}{\timex}\, \ekspon{-{\mathrm i} \omegaa  \tauoo} {\mathcal I}_{\ind{0}},\\ 
{\mathcal I}_{\ind{0}}= \int_{-\infty}^{\infty} & \frac{{\mathrm d}\dev }{2\pi}\,\ekspon{-{\mathrm i}\zeta_{\ind{0}} \dev }\, W(y) , \hspace{0.5cm}
\zeta_{\ind{0}} =  -\frac{\tauoo}{\timex}  + \frac{\cnsta}{2},\\
W(y)=  \cos & \left( \frac{\cnsta}{2} \sqrt{\dev^2+1} \right)+{\mathrm i}\dev\,\frac{ \sin\left(\frac{\cnsta}{2} \sqrt{\dev^2+1}\right) }{\sqrt{\dev^2+1}}, 
\end{split}
\end{equation}
and
\begin{equation}\label{lta032}
\begin{split}
G_{\ind{0H}}(\tauho)=&  -  \iu \, \frac{\cnstc\ \cnstd }{\timex}\, \ekspon{-{\mathrm i} \omegaa  \tauho} {\mathcal I}_{\ind{H}},\\ 
{\mathcal I}_{\ind{H}}=  \int_{-\infty}^{\infty} & \frac{{\mathrm d}\dev }{2\pi}\,\ekspon{-{\mathrm i}\zeta_{\ind{H}} \dev }\, V(y),\hspace{0.5cm}
\zeta_{\ind{H}} =  -\frac{\tauho}{\timex}   + \frac{\cnsta}{2},\\ 
V(y)= & \frac{\sin\left(\frac{\cnsta}{2}\sqrt{\dev^2+1}\right)}{\sqrt{\dev^2+1}}.
\end{split}
\end{equation}
The Fourier integral ${\mathcal I}_{\ind{H}}$ in \eqref{lta032} is a
tabulated integral \cite{GradshteynRyzhik3876} and can be calculated
analytically, as was previously carried out in solving similar problems
\cite{Kato68,Authier,MG03}:
\begin{equation}\label{lta042}
{\mathcal I}_{\ind{H}}=\frac{1}{2} J_{\ind{0}}\left(\sqrt{(\cnsta/2)^2-\zeta_{\ind{H}}^2}\right)
\Theta\left(\!\!\frac{\cnsta}{2}+\zeta_{\ind{H}}\!\!\right)
\Theta\left(\!\!\frac{\cnsta}{2}-\zeta_{\ind{H}}\!\!\right).
\end{equation}
Here $\Theta()$ is the Heaviside unit step function whose
value is zero for negative argument and one for positive argument.

The Fourier integral ${\mathcal I}_{\ind{0}}$ in \eqref{lta030} can be
calculated using the property ${\mathcal I}_{\ind{0}}=\partial
{\mathcal I}_{\ind{H}}/\partial (\cnsta/2) - \partial {\mathcal
  I}_{\ind{H}}/\partial \zeta $ \cite{Kato68,Authier,MG03} resulting in
\begin{widetext}
\begin{equation}\label{lta040}
{\mathcal I}_{\ind{0}}=\tilde{{\mathcal I}}_{\ind{0}}+  
2\delta\left(\!\!\frac{\cnsta}{2}-\zeta_{\ind{0}}\!\!\right),
\hspace{0.5cm}  
\tilde{{\mathcal I}}_{\ind{0}}=-\frac{\cnsta/2 +\zeta_{\ind{0}}}{2\sqrt{(\cnsta/2)^2-\zeta_{\ind{0}}^2}}J_{\ind{1}}\left(\sqrt{(\cnsta/2)^2-\zeta_{\ind{0}}^2}\right)
\Theta\left(\!\!\frac{\cnsta}{2}+\zeta_{\ind{0}}\!\!\right)
\Theta\left(\!\!\frac{\cnsta}{2}-\zeta_{\ind{0}}\!\!\right).
\end{equation}
Here the $\delta$-function appears as a result of differentiating the
step functions.  Finally with \eqref{lta042} and \eqref{lta040} and
definition of $\zeta_{\ind{0}}$ and $\zeta_{\ind{H}}$ in
\eqref{lta030}-\eqref{lta032} we arrive at the following analytical
expressions for the plane-wave response functions in Laue-case
geometry
\begin{equation}\label{lta051}
\begin{split}
  G_{\ind{00}}(\tauoo)\,=\,&\tilde{G}_{\ind{00}}(\tauoo)\,+\,
\cnstc\, \delta ( \tauoo ),\\ 
\tilde{G}_{\ind{00}}(\tauoo)\,=\,&\frac{\cnstc}{2\timexo} \ekspon{-{\mathrm i}
    \omegaa \tauoo}\,
 \left(1-\frac{\tauoo}{\thicknessduration } \right) \frac{J_{\ind{1}}\left[\sqrt{\frac{\tauoo}{\timexo} \left(1-\frac{\tauoo}{\thicknessduration}\right)}\, \right]}{\sqrt{\frac{\tauoo}{\timexo} \left(1-\frac{\tauoo}{\thicknessduration}\right)}}\, 
\Theta( \tauoo )\Theta\left( \thicknessduration -\tauoo \right)  ,
\end{split}
\end{equation}
\begin{equation}
  G_{\ind{0H}}(\tauho)\,=\, -  \iu \,\frac{\cnstc\ \cnstd}{2 \timex }\,
  \ekspon{-{\mathrm i} \omegaa \tauho}\,
  J_{\ind{0}}\left[\sqrt{\frac{\tauho}{\timexo } \left(1-\frac{\tauho}{\thicknessduration }\right)}\,  \right]
\Theta( \tauho )\Theta\left( \thicknessduration -\tauho \right)  .
\label{lta053}
\end{equation}

\end{widetext}

With one exception, these expressions agree with relevant expressions
obtained by Malgrange and Graeff in \cite{MG03}, where diffraction of
short x-ray pulses with infinite wavefront in the asymmetric Laue-case
was studied analytically. Unlike the expression for forward
diffraction presented in \cite{MG03}, Eq.~\eqref{lta051} contains the
delta-function, which represents the prompt response in the forward
diffraction due to spectral components far from the Bragg diffraction
region that propagate essentially diffraction-free through the
crystal.

We note also, that a reference system $(x^{\prime},z^{\prime})$ was
used in \cite{MG03} attached to the crystal rear surface. Unlike this,
we are using in our treatment for each diffracted wavefield its own
reference system $(\ucvcx,\wcvcx) $. We are also using a different
approach to calculate the vacuum wavevector of the diffracted wave
\eqref{pro027}-\eqref{pro0333}.  Due to these differences, the
expressions for the spatiotemporal variables $\tauxo$
\eqref{pro024}-\eqref{pro028}, and similar variables in \cite{MG03} -
Eq.~(28) - may appear at a first glance to be very different. However,
our detailed comparison shows that they are actually identical. So
mathematically our results and results of paper \cite{MG03} for the
delayed parts of the response functions are in agreement, except for
the delta function in Eq.~\eqref{lta051}.

\section{Mapping Time  on Lateral Space Shift}
\label{mapping}

\begin{figure*}[t!]
\setlength{\unitlength}{\textwidth}
\begin{picture}(1,0.51)(0,0)
\put(0.0,0.00){\includegraphics[width=0.49\textwidth]{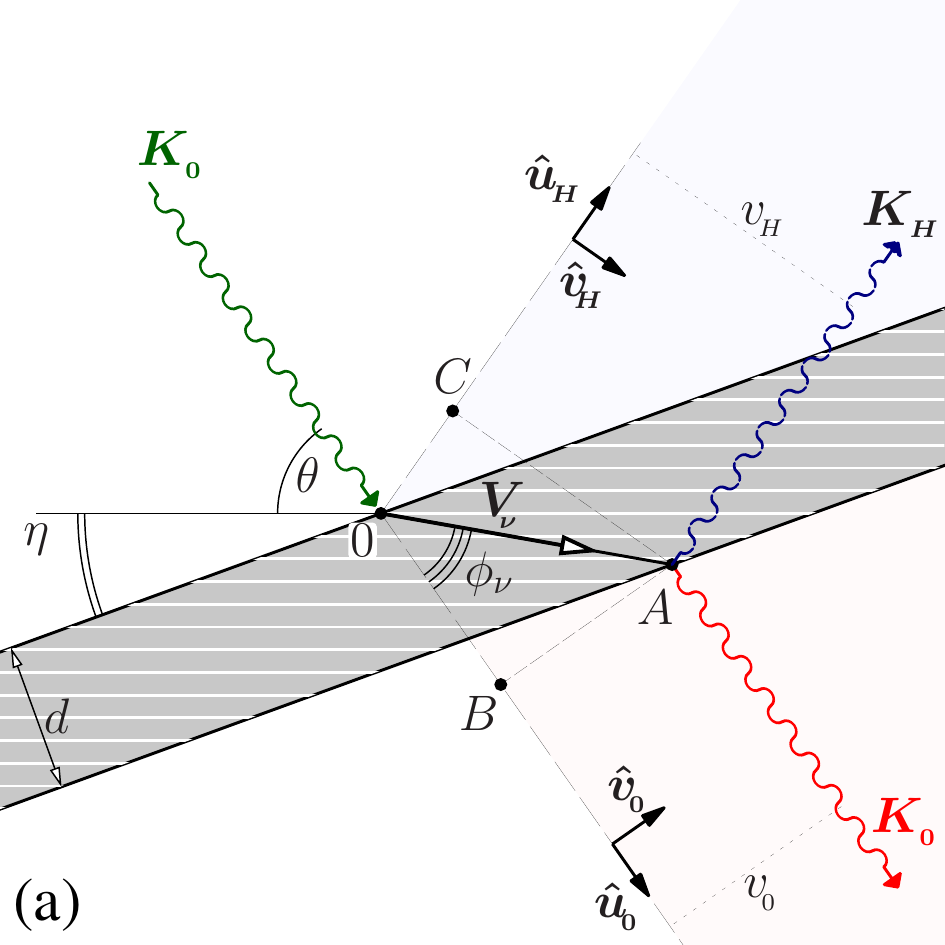}}
\put(0.50,0.00){\includegraphics[width=0.49\textwidth]{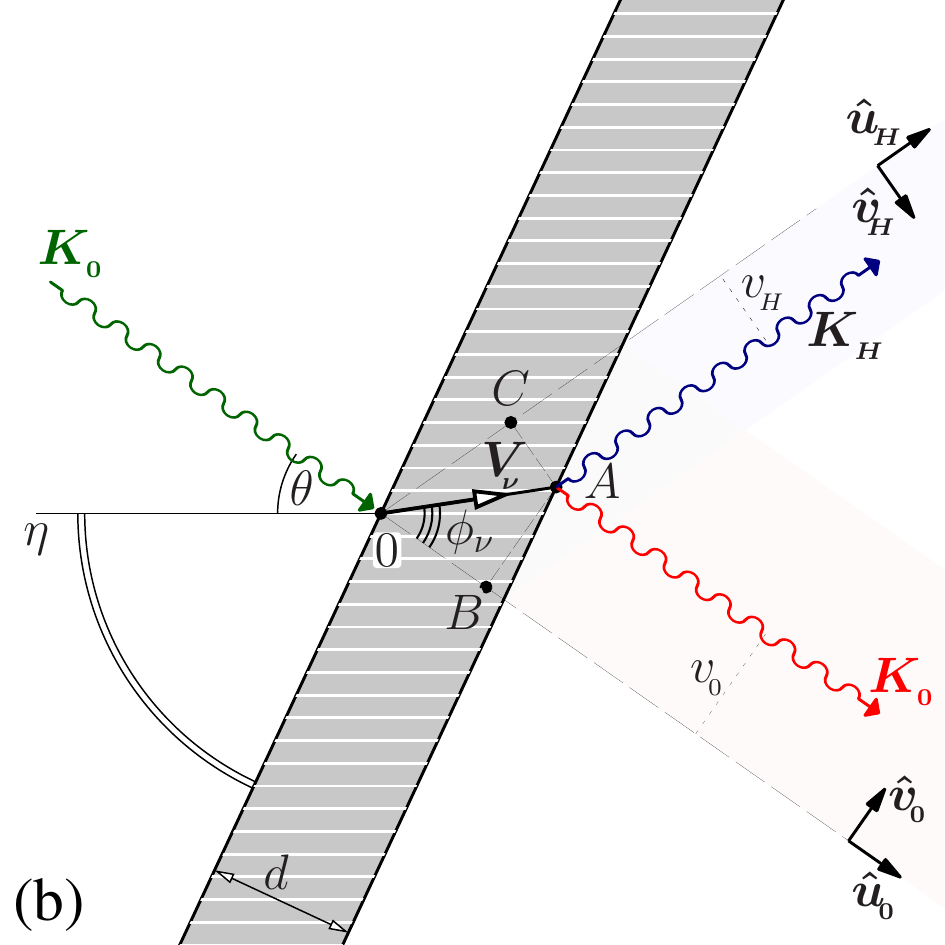}}
\end{picture}
\caption{Schematic presentation of two-beam x-ray Bragg diffraction
  from a crystal (a) in the reflection (Bragg) scattering geometry and
  (b) in the transmission (Laue) scattering geometry.  The direction
  of the energy flow by the crystal wavefield is indicated by the
  vector of the group velocity $\gvvec$, at an angle $\phi_{\nu}$ to
  the direction of the incident wave. }
\label{fig005}
\end{figure*}

The relationship $\wcx = \taux c \cot\thetac$ between the time delay
$\taux$ and the spatial shift $\wcx$ in Bragg diffraction,
representing the trace of the Bragg's law dispersion envelope
$\Pi(\wcx - \taux c \cot\thetac)$ in Eq.~\eqref{pro060}, can be
derived alternatively by combining Ewald's concept of the crystal
wavefield \cite{Ewald17} with the concept of energy flow introduced by
von~Laue~\cite{Laue52}.

These concepts lead to the following picture of physical processes
involved in x-ray Bragg diffraction in crystals. It is illustrated
graphically in Figs.~\ref{fig005}(a) and (b), schematically presenting
diffraction in the Bragg-case geometry and in the Laue-case geometry,
respectively.

An incident monochromatic plane wave with wavevector $\kovc$ excites
monochromatic wavefields in the crystal given by
Eq.~\eqref{wavefield}, where each wavefield is associated with a tie
point on one of the brunches of the dispersion surface, which we
number below by $\nu$.  The energy flow for each wavefield in a
perfect crystal is given by the wavefield Poynting vector
\cite{Laue52}, which is directed along the normal to the dispersion
surface taken at the tie point of the surface representing the field
\cite{Kato58}.  The Poynting vector is parallel to the vector of group
velocity $\gvvec$.  As a result, different monochromatic wavefields
propagate, first, along different paths of different lengths, and
second, with different group velocities $\gvvec$.  At the exit
surface, the wavefield breaks up into independent plane wavefields,
one with the wavevector $\kovc$ propagating in the direction $\ucvco$,
and another with the wavevector $\khvc$ propagating in the direction
$\ucvch$. One should note that the concept of the energy flow works
with one substantial limitation: it is not applicable in the total
reflection region in Bragg-case geometry, as there is no propagating
through the crystal wavefields in this case.  It is applicable,
however, outside the total reflection region. Due to this, the wave
with the wavevector $\khvc$ is shown in Fig.~\ref{fig005}(a)
propagating along the line starting at point $A$.

Using Figs.~\ref{fig005}(a) and (b) we calculate for the spatial
shifts $\wcx$ and delays $\taux$:
\begin{gather}\label{wcowch}
\wco\equiv AB=OA\,\sin\phi_{\nu}, \hspace{0.25cm} \tauo=\frac{OA}{\gvval}-\frac{OB}{c},\\
\wch\equiv AC=OA\,\sin(2\theta-\phi_{\nu}), \hspace{0.25cm} \tauh=\frac{OA}{\gvval}-\frac{OC}{c}. 
\end{gather}
The magnitude of the group velocity can be given, as derived in
Appendix~\eqref{wg-velocity-angle}, by
\begin{equation}
\frac{\gvval}{c} \,=  \, \frac{\cos\theta}{\cos(\theta-\phi_{\nu})}.
\label{eq003-2}
\end{equation} 
We use in Eqs.~\eqref{wcowch} the fact that propagation along $OB$
($\phi_{\nu}=0$) or $OC$ ($\phi_{\nu}=2\theta$) takes place with the
speed of light in vacuum $c$, in agreement with Eq.~\eqref{eq003-2}.
Using the relationships $OB=OA\,\cos\phi_{\nu}$,
$OC=OA\cos(2\theta-\phi_{\nu})$, and Eq.~\eqref{eq003-2}, we obtain
\begin{equation}
\wcx\,=\,\taux c \,\cot\theta,\hspace{0.5cm} \hvar=(0,H).
\label{eq004-2}
\end{equation} 
This relationship is valid in the general case of asymmetric
diffraction, both for Bragg and Laue scattering geometries.  It maps
temporal onto spatial scales in Bragg diffraction, in agreement with
the Bragg's law dispersion envelope $\Pi(\wcx - \taux c \cot\thetac)$
in Eq.~\eqref{pro060}.  Uncertainty relationships are always valid,
and therefore Eq.~\eqref{eq004-2} are actually applied not for
absolutely monochromatic waves, or waves localized in time and space,
but rather for wave packets with certain spectral and momentum
distributions.

\section{Wavefield Group Velocity in the Crystal vs Propagation Angle}
\label{wg-velocity-angle}  

The group velocity vector $\gvvec$ is given by
\cite{Laue52,Kato58,Authier}:
\begin{equation}
\gvvec\, =\, c \frac{\ucvco  + \ucvch R_{\nu}^2}{ 1 + R_{\nu}^2},
\label{we003xx}
\end{equation}
where $R_{\nu}$ is defined in Eq.~\eqref{rta020}, and $\nu$ numbers
brunches of the dispersion surface.

The absolute value of the group velocity $\gvval$ can be calculated by
taking the magnitude of Eq.~\eqref{we003xx} and recalling that $\ucvco
\, \ucvch = \cos 2\theta$:
\begin{equation}
\gvval\, =\, c \frac{\sqrt{1  + 2  R_{\nu}^2 \cos\2\theta +  R_{\nu}^4 }}{ 1 + R_{\nu}^2}
\label{app020}
\end{equation}
The angle $\phi_{\nu}$ between the direction of the group velocity
vector $\gvvec$ and optical axis $\ucvco$ is determined from the vector scalar product
$\cos\phi_{\nu}  = \ucvco\gvvec/\gvval$, which using  Eq.~\eqref{we003xx} becomes
\begin{equation}
\cos\phi_{\nu} \, =\, \frac{ 1 + R_{\nu}^2 \cos\2\theta}{\sqrt{1  + 2  R_{\nu}^2 \cos\2\theta +  R_{\nu}^4 }}.
\label{app030}
\end{equation}
Combining \eqref{app020}-\eqref{app030} we obtain the following relationship between
the magnitude of wavefield group velocity $\gvval$ and its direction $\phi_{\nu}$   
\begin{equation}
\frac{\gvval}{c} \,=  \, \frac{\cos\theta}{\cos(\theta-\phi)},
\label{app100}
\end{equation} 
Equation~\eqref{app100} gives a physically reasonable result. The
direction of the wavefield propagation under the Bragg diffraction
condition is at $\phi=\theta$, in which case the group velocity
$\gvval \,= \,c\, \cos\theta $ is less than speed of light in
vacuum. Far from Bragg diffraction conditions $\phi=0$ or
$\phi=2\theta$, resulting in a reasonable solution $\gvval \,= \,c $.

\begin{widetext} 
\section{Bragg Reflections in      Diamond} 
\label{diamondreflections} 
\begin{table}[h!]
      \centering 
\hspace*{0.0\linewidth} 
\begin{minipage}[t]{0.30\linewidth} 
\begin{tabular}{|r|rrrr|} 
\hline
      $h~k~l$     &   $\ebr$      &  $\extlengs$ & $\whhs$        &  $\Deltaeh$ \\[-0.0pt]
      &         &   &         &   \\[-0.0pt]
      &    [keV]     & [$\mu$m]   &  $\times 10^{-5}$       &  [meV]  \\[-0.0pt]
      &         &   &         &   \\[-0.0pt]
      \hline
      1    1    1 &   3.01034  &  1.09  &  8.17  &  192. \\[-0.0pt]
      2    2    0 &   4.91561  &  1.98  &  3.04  &  106. \\[-0.0pt]
      3    1    1 &   5.76401  &  3.74  &  2.20  &  56.0 \\[-0.0pt]
      4    0    0 &   6.95161  &  3.63  &  1.51  &  60.6 \\[-0.0pt]
      3    3    1 &   7.57532  &  5.89  &  1.27  &  35.8 \\[-0.0pt]
      4    2    2 &   8.51391  &  5.03  &  1.00  &  44.5 \\[-0.0pt]
      3    3    3 &   9.03035  &  7.83  &  0.89  &  27.3 \\[-0.0pt]
      5    1    1 &   9.03035  &  7.83  &  0.89  &  27.3 \\[-0.0pt]
      4    4    0 &   9.83108  &  6.41  &  0.75  &  35.9 \\[-0.0pt]
      5    3    1 &  10.2815  &  9.82  &  0.69  &  22.6 \\[-0.0pt]
      6    2    0 &  10.9914  &  7.87  &  0.60  &  29.2 \\[-0.0pt]
      5    3    3 &  11.3961  &  11.9  &  0.56  &  19.1 \\[-0.0pt]
      4    4    4 &  12.0404  &  9.44  &  0.50  &  25.0 \\[-0.0pt]
      5    1    5 &  12.4110  &  14.2  &  0.47  &  16.4 \\[-0.0pt]
      7    1    1 &  12.4110  &  14.2  &  0.47  &  16.4 \\[-0.0pt]

      \hline \end{tabular}
  \end{minipage} 
\hspace{0.01\linewidth} 
\begin{minipage}[t]{0.3\linewidth} %
\centering %
\begin{tabular}{|r|rrrr|} 
\hline
  $h~k~l$     &   $\ebr$      &  $\extlengs$ & $\whhs$        &  $\Deltaeh$ \\[-0.0pt]
  &         &   &         &   \\[-0.0pt]
  &    [keV]     & [$\mu$m]   &  $\times 10^{-5}$       &  [meV]  \\[-0.0pt]
  &         &   &         &   \\[-0.0pt]
  \hline
  6    4    2 &  13.0051  &  11.1  &  0.43  &  21.3 \\[-0.0pt]
  5    5    3 &  13.3489  &  16.7  &  0.40  &  14.3 \\[-0.0pt]
  7    3    1 &  13.3489  &  16.7  &  0.40  &  14.3 \\[-0.0pt]
  8    0    0 &  13.9030  &  13.0  &  0.37  &  18.6 \\[-0.0pt]
  7    3    3 &  14.2251  &  19.5  &  0.36  &  12.6 \\[-0.0pt]
  6    6    0 &  14.7464  &  15.1  &  0.33  &  16.5 \\[-0.0pt]
  8    2    2 &  14.7464  &  15.1  &  0.335  &  16.5 \\[-0.0pt]
  7    5    1 &  15.0504  &  22.5  &  0.322  &  10.8 \\[-0.0pt]
  8    4    0 &  15.5440  &  17.3  &  0.301  &  14.7 \\[-0.0pt]
  7    5    3 &  15.8328  &  25.7  &  0.291  &  9.9 \\[-0.0pt]
  9    1    1 &  15.8328  &  25.7  &  0.291  &  9.9 \\[-0.0pt]
  6    6    4 &  16.3027  &  19.7  &  0.274  &  12.9 \\[-0.0pt]
  9    3    1 &  16.5783  &  29.2  &  0.265  &  8.5 \\[-0.0pt]
  8    4    4 &  17.0276  &  22.3  &  0.251  &  11.6 \\[-0.0pt]
  7    5    5 &  17.2916  &  33.0  &  0.243  &  7.8 \\[-0.0pt]

  \hline \end{tabular}
\end{minipage} %
\hspace{0.01\linewidth} \begin{minipage}[t]{0.3\linewidth} %
\centering %
\begin{tabular}{|r|rrrr|} 
\hline
$h~k~l$     &   $\ebr$      &  $\extlengs$ & $\whhs$        &  $\Deltaeh$ \\[-0.0pt]
&         &   &         &   \\[-0.0pt]
&    [keV]     & [$\mu$m]   &  $\times 10^{-5}$       &  [meV]  \\[-0.0pt]
&         &   &         &   \\[-0.0pt]
\hline
7    7    1 &  17.2916  &  33.0  &  0.243  &  7.8 \\[-0.0pt]
9    3    3 &  17.2916  &  33.0  &  0.243  &  7.8 \\[-0.0pt]
10    0    2 &  17.7229  &  25.1  &  0.232  &  10.1 \\[-0.0pt]
8    6    2 &  17.7229  &  25.1  &  0.232  &  10.1 \\[-0.0pt]
7    7    3 &  17.9767  &  37.2  &  0.225  &  6.9 \\[-0.0pt]
9    5    1 &  17.9767  &  37.2  &  0.225  &  6.9 \\[-0.0pt]
9    5    3 &  18.6366  &  41.6  &  0.210  &  6.1 \\[-0.0pt]
10    4    2 &  19.0375  &  31.5  &  0.201  &  8.1 \\[-0.0pt]
11    1    1 &  19.2740  &  46.4  &  0.196  &  5.4 \\[-0.0pt]
7    7    5 &  19.2740  &  46.4  &  0.196  &  5.4 \\[-0.0pt]
8    8    0 &  19.6618  &  35.1  &  0.188  &  7.2 \\[-0.0pt]
11    3    1 &  19.8909  &  51.6  &  0.184  &  5.0 \\[-0.0pt]
9    5    5 &  19.8909  &  51.6  &  0.184  &  5.0 \\[-0.0pt]
9    7    1 &  19.8909  &  51.6  &  0.184  &  5.0 \\[-0.0pt]
&           &        &         &      \\[-0.0pt]
\hline    
\end{tabular} 
\end{minipage} 
\caption{ Allowed Bragg reflections
  $hkl$ in diamond crystals and their parameters relevant to the present
  studies: Bragg energy $\ebr\,=\,hc/2\dhkl$, the extinction length
  $\extlengs$ \eqref{extleng}, the Bragg's law correction $\whhs$
  \eqref{rta074}, the energy width $\Deltaeh$ (at $\theta=\pi/2$). These
  parameters are calculated using interplanar distance
  $\dhkl=3.56712(2)$~\AA\ in diamond crystals at $T=298$~K
  \cite{SSh10,SSh11}, a Debye Temperature of 2230~T
  \cite{Gschneidner64}, and anomalous scattering factors from
  \cite{KP90,KP95}.}

\label{tab1}
\end{table}
\end{widetext}


\end{document}